\documentclass[journal]{IEEEtran}

\ifCLASSINFOpdf

\else

\fi
\usepackage[cmex10]{amsmath}

\usepackage{amssymb}

\usepackage{graphics,graphicx,epic,eepic,epsfig,times,units} % epstopdf

\newcommand{\reals}{{\mbox{\bf R}}}

\usepackage{float}

\usepackage{psfrag}

\usepackage{ifpdf}

\usepackage{cite}

\usepackage{siunitx}

\usepackage{multirow}

\usepackage{xcolor}

\usepackage{algorithm,algorithmicx}
\usepackage{algpseudocode}
\usepackage{algcompatible}
\algrenewcommand\alglinenumber[1]{\scriptsize #1:}
\newcommand{\algrule}[1][.2pt]{\par\vskip.5\baselineskip\hrule height #1\par\vskip.5\baselineskip}
\usepackage{tabularx}

\begin{document}

\title{A Fast Distributed Algorithm for Large-Scale Demand Response Aggregation}

\author{Sleiman~Mhanna,~\IEEEmembership{Student~MIEEE,}
        Archie~C.~Chapman,~\IEEEmembership{~MIEEE}
        and~Gregor~Verbi\v{c},~\IEEEmembership{Senior~MIEEE,}
%\thanks{Sleiman~Mhanna, Gregor~Verbi\v{c}~and~Archie~C.~Chapman are with the School of Electrical and Information Engineering, The University of Sydney, Australia, (e-mail: sleiman.mhanna@sydney.edu.au; gregor.verbic@sydney.edu.au; archie.chapman@sydney.edu.au).}
%\thanks{Manuscript received December 19, 2014; revised December 27, 2015.}
}

%\markboth{IEEE TRANSACTIONS ON SMART GRID,~Vol.~3, No.~3, February~2015}
%{Mhanna \MakeLowercase{\textit{et al.}}: A Distributed Algorithm for Large-Scale Demand Response Aggregation}

\maketitle

\begin{abstract}
A major challenge to implementing residential demand response is that of aligning the objectives of many households, each of which aims to minimize its payments and maximize its comfort level, while balancing this with the objectives of an aggregator that aims to minimize the cost of electricity purchased in a pooled wholesale market. This paper presents a fast distributed algorithm for aggregating a large number of households with a mixture of discrete and continuous energy levels. A distinctive feature of the method in this paper is that the nonconvex DR problem is decomposed in terms of households as opposed to devices, which allows incorporating more intricate couplings between energy storage devices, appliances and distributed energy resources. The proposed method is a fast distributed algorithm applied to the double smoothed dual function of the adopted DR model. The method is tested on systems with up to $2560$ households, each with $10$ devices on average. The proposed algorithm is designed to terminate in $60$ iterations irrespective of system size, which can be ideal for an on-line version of this problem. Moreover, numerical results show that with minimal parameter tuning, the algorithm exhibits a very similar convergence behavior throughout the studied systems and converges to near-optimal solutions, which corroborates its scalability.
\end{abstract}

\begin{IEEEkeywords}
Dual decomposition, accelerated gradient methods, demand response aggregation, smoothing techniques, mixed-integer variables, smart grid, energy management.
\end{IEEEkeywords}

\IEEEpeerreviewmaketitle

\section*{Notation}
%\addcontentsline{toc}{section}{Notation}
\subsection{Acronyms}
\begin{IEEEdescription}[\IEEEsetlabelwidth{$T_{i}^{\mathrm{comf,n}}$}\IEEEusemathlabelsep]
\item[CHP] Combined heat and power.
\item[DR] Demand Response.
\item[HVAC] Heating, ventilation and air conditioning.
\item[MIP] Mixed-integer program/programming.
\item[MIQP] Mixed-integer quadratic program/programming.
\item[MINLP] Mixed-integer nonlinear program/programming.
\item[QP] Quadratic program/programming.
\item[SoC] State of charge.
\end{IEEEdescription}

\subsection{Parameters}
%\small
\begin{IEEEdescription}[\IEEEsetlabelwidth{$T_{i}^{\mathrm{comf,n}}$}\IEEEusemathlabelsep]
\item[$A_{i}$] Agent $i$'s total number of devices.
\item[$C^{t}\left(x_{g}^{t}\right)$] Electricity cost ($\SI{}{\$}$) of drawing $x_{g}^{t}$ units of energy from the grid during time-slot $t$.
\item[$c0^{t}$] Coefficient ($\SI{}{\$}$) of the constant term in $C^{t}\left(x_{g}^{t}\right)$ during time-slot $t$.
\item[$c1^{t}$] Coefficient ($\SI{}{\$/\kilo\watt\hour}$) of the linear term in $C^{t}\left(x_{g}^{t}\right)$ during time-slot $t$.
\item[$c2^{t}$] Coefficient ($\SI{}{\$/\kilo\watt\hour^2}$) of the quadratic term in $C^{t}\left(x_{g}^{t}\right)$ during time-slot $t$.
\item[$D^{t}_{i,a}\left(\cdot\right)$] Dissatisfaction cost ($\SI{}{\$}$) incurred by agent $i$'s Type 2, Type 3 or Type 6 device $a$ during time-slot $t$.
\item[$e^{l}_{i,a}$] Energy ($\SI{}{\kilo\watt\hour}$) consumed by agent $i$'s Type 1, Type 2 or Type 3 device $a$ during time-step $\Delta \tau$.
\item[$e^{\mathrm{SoC,ini}}_{i,a}$] Initial state of energy ($\SI{}{\kilo\watt\hour}$) of agent $i$'s Type 4 or Type 5 device $a$.
\item[$e^{\mathrm{SoC,final}}_{i,a}$] Final state of energy ($\SI{}{\kilo\watt\hour}$) of agent $i$'s Type 4 or Type 5 device $a$.
\item[$e^{\mathrm{SoC,min}}_{i,a}$] Minimum state of energy ($\SI{}{\kilo\watt\hour}$) of agent $i$'s Type 4 or Type 5 device $a$.
\item[$e^{\mathrm{SoC,max}}_{i,a}$] Maximum state of energy ($\SI{}{\kilo\watt\hour}$) of agent $i$'s Type 4 or Type 5 device $a$.
\item[$E_{i,a}$] Total energy ($\SI{}{\kilo\watt\hour}$) requirement over $\mathcal{T}$ of agent $i$'s Type 3 device $a$. 
\item[$G^{\mathrm{max}}$] Maximum power ($\SI{}{\kilo\watt}$) that can be drawn from the grid.
\item[$I$] Total number of household agents.
\item[$\eta^{\mathrm{ch}}_{i,a}$] Charging efficiency of agent $i$'s Type 4 or Type 5 device $a$.
\item[$\eta^{\mathrm{dis}}_{i,a}$] Discharging efficiency of agent $i$'s Type 4 or Type 5 device $a$.
\item[$k$] Iteration number.
\item[$l$] Operating mode of agent $i$'s Type 1, Type 2 or Type 3 device $a$.
\item[$L$] Total number of operating modes of agent $i$'s Type 1, Type 2 or Type 3 device $a$.
\item[$p^{l}_{i,a}$] Power level ($\SI{}{\kilo\watt}$) at operation mode $l$ of agent $i$'s Type 1, Type 2 or Type 3 device $a$.
\item[$P^\mathrm{ch,min}_{i,a}$] Minimum charging power ($\SI{}{\kilo\watt}$) of agent $i$'s Type 4 or Type 5 device $a$.
\item[$P^\mathrm{ch,max}_{i,a}$] Maximum charging power ($\SI{}{\kilo\watt}$) of agent $i$'s Type 4 or Type 5 device $a$.
\item[$P^\mathrm{dis,min}_{i,a}$] Minimum discharging power ($\SI{}{\kilo\watt}$) of agent $i$'s Type 4 or Type 5 device $a$.
\item[$P^\mathrm{dis,max}_{i,a}$] Maximum discharging power ($\SI{}{\kilo\watt}$) of agent $i$'s Type 4 or Type 5 device $a$.
\item[$P_{i}^{\mathrm{PV},t}$] Predicted power ($\SI{}{\kilo\watt}$) generation of agent $i$'s PV system at time-slot $t$.
\item[$P^\mathrm{th,min}_{i,a}$] Minimum power requirement ($\SI{}{\kilo\watt}$) of agent $i$'s Type 6 device $a$.
\item[$P^\mathrm{th,max}_{i,a}$] Maximum power requirement ($\SI{}{\kilo\watt}$) of agent $i$'s Type 6 device $a$.
\item[$P^\mathrm{max}_{i}$] Maximum power rating ($\SI{}{\kilo\watt}$) of the household main circuit breaker's overload protection.
\item[$\phi$] Type of agent $i$'s device $a$.
\item[$\pi_{i,a}$] Minimum `on' time of a Type 3 device $a$.
\item[$\Delta \tau$] Time resolution ($\SI{0.25}{\hour}$ or $\SI{1}{\hour}$).
\item[$T$] Length of the decision time horizon.
\item[$T_{i}^{\mathrm{in,min}}$] Minimum temperature value ($\SI{}{\degreeCelsius}$) in agent $i$'s comfortable temperature range.
\item[$T_{i}^{\mathrm{in,max}}$] Maximum temperature value ($\SI{}{\degreeCelsius}$) in agent $i$'s comfortable temperature range.
\item[$T_{i}^{\mathrm{comf}}$] Agent $i$'s most comfortable temperature ($\SI{}{\degreeCelsius}$).
\item[$T^{t}_{\mathrm{out}}$] Outside temperature ($\SI{}{\degreeCelsius}$) at time-slot $t$.
\item[$\tau_{i,a}$] Agent $i$ device $a$'s desired scheduling interval.
\item[$\tau^\mathrm{start}_{i,a}$] Start time of agent $i$ device $a$'s desired scheduling interval.
\item[$\tau^\mathrm{end}_{i,a}$] End time of agent $i$ device $a$'s desired scheduling interval.
\item[$\gamma^{l}_{i,a}$] Nonnegative parameter ($\SI{}{\$}$) that reflects agent $i$'s preference for operating mode $l$ of Type 2 appliance $a$.
\item[$\overline{\gamma}_{i,a}$,$\underline{\gamma}_{i,a}$] Nonnegative parameters ($\SI{}{\$}$) that determine how quickly the user gets dissatisfied when the scheduled operation of Type 3 device $a$ is delayed by $t-(\tau^{\mathrm{end}}_{i,a}+\pi_{i,a}-\Delta \tau)$ time-slots away from $\tau^{\mathrm{end}}_{i,a}$ or advanced $\tau^{\mathrm{start}}_{i,a}-t$ time-slots ahead of $\tau^{\mathrm{start}}_{i,a}$, respectively.
\item[$\gamma^{\mathrm{th}}_{i,a}$] Nonnegative parameter ($\SI{}{\$/\degreeCelsius^2}$) that depends on agent $i$'s tolerance to deviations of the inside temperature $T_{i}^{\mathrm{in},t}$ from agent $i$'s most comfortable temperature $T_{i}^{\mathrm{comf}}$.
\item[$\psi_{i,a}$] Parameter ($\SI{}{\degreeCelsius/\kilo\watt\hour}$) of the thermal dynamics equation.
\item[$\zeta_{i,a}$] Parameter of the thermal dynamics equation.
\item[$\mu$] Smoothness parameter ($\SI{}{\$/\kilo\watt\hour^2}$).
\item[$\nu$] Penalty parameter ($\SI{}{\$/\kilo\watt\hour^2}$).
\item[$\kappa$] Strong concavity parameter ($\SI{}{\kilo\watt\hour^2/\$}$).
\end{IEEEdescription}

\subsection{Sets}
\begin{IEEEdescription}[\IEEEsetlabelwidth{$T_{i}^{\mathrm{comf,n}}$}\IEEEusemathlabelsep]
\item[$\mathcal{A}_{i,\phi}$] Set of agent $i$'s type $\phi$ devices.
\item[$\mathcal{A}_{i}$] Set of all devices of agent $i$.
\item[$\mathcal{I}$] Set of all household agents.
\item[$\mathcal{T}$] DR decision time horizon.
\end{IEEEdescription}

\subsection{Variables}
\begin{IEEEdescription}[\IEEEsetlabelwidth{$T_{i}^{\mathrm{comf,n}}$}\IEEEusemathlabelsep]
\item[$T_{i}^{\mathrm{in},t}$] Inside temperature ($\SI{}{\degreeCelsius}$) at time-slot $t$.
\item[$u^{l,t}_{i,a}$] Binary variable that takes a value of `1' when agent $i$'s Type 2 or Type 3 device $a$ is in operating mode $l$ during time-slot $t$.
\item[$u^{\mathrm{ch},t}_{i,a}$] Binary variable that takes a value of `1' when agent $i$'s Type 4 or Type 5 device $a$ is in charging mode during time-slot $t$.
\item[$u^{\mathrm{dis},t}_{i,a}$] Binary variable that takes a value of `1' when agent $i$'s Type 4 or Type 5 device $a$ is in discharging mode during time-slot $t$.
\item[$v^{t}_{i,a}$] Startup binary variable of agent $i$'s Type 3 device $a$ during time-slot $t$.
\item[$x^{\mathrm{ch},t}_{i,a}$] Charging energy ($\SI{}{\kilo\watt\hour}$) of agent $i$'s Type 4 or Type 5 device $a$ during time-slot $t$.
\item[$x^{\mathrm{dis},t}_{i,a}$] Discharging energy ($\SI{}{\kilo\watt\hour}$) of agent $i$'s Type 4 or Type 5 device $a$ during time-slot $t$.
\item[$x^{\mathrm{SoC},t}_{i,a}$] State of energy ($\SI{}{\kilo\watt\hour}$) of agent $i$'s Type 4 or Type 5 device $a$ at time-slot $t$.
\item[$x_{i,a}^{t}$] Energy consumption ($\SI{}{\kilo\watt\hour}$) of agent $i$'s device $a$ during time-slot $t$.
\item[$x_{0}^{t}$] Total energy demand ($\SI{}{\kilo\watt\hour}$) during time-slot $t$.
\item[$x_{g}^{t}$] Energy ($\SI{}{\kilo\watt\hour}$) drawn from the grid during time-slot $t$.
\item[$\boldsymbol{\lambda}$] Vector of Lagrange multipliers ($\SI{}{\$/\kilo\watt\hour}$). 
\end{IEEEdescription}

\section{Introduction}

\IEEEPARstart{D}{emand} response programs capitalize on advancements in communications, control, and computation technologies of the future grid to harness the flexibility of electric loads for demand shaping, supply-demand balancing and other ancillary services. Central to the vision of the future grid is the deployment of smart meters with embedded agents that represent the consumers in their interaction with a DR aggregator. This technology can enable efficient participation of flexible loads in energy markets through leveraging carefully designed price and load information exchange schemes.
 
Given this context, efficient load scheduling and aggregation is a problem of growing importance in the area of demand response. However, the problem of scheduling large numbers of household loads, which comprise 25-30\% of system load in advanced economies and higher elsewhere, is particularly challenging for two main reasons. First, household agents are self-interested and aim at minimizing their costs and maximizing their comfort levels, whereas the aggregator aims at decreasing peak demand and minimizing the cost of electricity purchased in a pooled wholesale market. Therefore, the challenge lies in devising a coordination scheme to aggregate these households into a usable DR resource that aligns the objectives of the households with the objectives of the aggregator.

Second, many household electrical devices have discrete operating points that can only be represented by mixed-integer variables (as in \cite{User-sensitiveScheduling,schedulingwithMIC,SHEMSwithcomfortablelifestyle,RTPbasedDR,RDRwithMulticlassApps,RTDRStackelberg,FaithfulMDinDR,MILPforDSM,JointOptofEVandHEMS,JointSchedulingEVandDR,EMtwohorizonalgorithm}), and some household device uses are often coupled, thus giving household electricity demand a combinatorial structure \cite{ahealthydoseofreality}. However, most energy management methods, such as those in \cite{User-sensitiveScheduling, schedulingwithMIC,SHEMSwithcomfortablelifestyle,RTPbasedDR,EMtwohorizonalgorithm,RDRwithMulticlassApps,RTDRStackelberg,algorithmforHEMSandDR,DynamicDRCbasedonRTP,EMSandDynamicprice,HEMSwithsolarTLandRE,ManagingDERs,Althaher_ADRfromHEMS}, address only one facet of the DR problem, which is local energy and comfort management. That is, they do not address system-wide aggregation of these DR capable households; moreover, the methods proposed in these works are either incompatible with wide-area aggregation or simply intractable in large-scale problems \cite{ChapmanVH_IntegratingDSA}.

The presence of mixed-integer variables results in a mixed-integer program (MIP) that has a NP-hard computational complexity. Therefore, solving the DR aggregation problem centrally, as in \cite{MILPforDSM,FaithfulMDinDR,JointOptofEVandHEMS,JointSchedulingEVandDR,Igualada_EMSwithV2G}, may spell intractability when the number of households is large. Furthermore, solving this problem centrally requires sending all of the households' private information to the aggregator, which entails substantial communication overhead and privacy concerns.
 
To this end, distributed methods are emerging as a way of efficiently implementing large-scale DR. The existing literature on distributed methods for demand response is split into two main categories. The first category includes methods that treat the household energy levels as continuous \cite{NaLi,PARconstrainedDR,UserCentricDR,optimalRTP,DependableDR,RLCwithlostAMIs,EFTPDSM,Disaggregatedbundle,decompositionformarketclearingwithDR,DRinLargepopulation,RLScostefficiency,DistributedDRwithEVs,IncrementalWelfareconsensus,RDRwithRTPinMAS,randomizedADMM,DRusingDantzigWolfe,supplyfunctionbidding,TchebycheffDecomposition,Atzeni_DSMviaDERandStorage,Atzeni_NoncooperativeandCooperativeDSM}, which often renders the underlying DR problem convex and therefore computationally conducive. The second category of papers includes the more realistic methods that treat the household energy levels as a mixture of discrete and continuous and account for inter-temporal device couplings \cite{ElectricitymarketsandDR,EMSwithDR,Zheng_DRusingLyapunov,DistributedAlgorithmforRDR,OptimalDecentralizedRDR,RDRwithinterruptibletasks,distributedalgorithmforHEMS,DRwithMIC,Yaagoubi_userawareGT,Tushar_distriburtedRTmechanism,onlineauctionforDR,distributedalgorithmMIP}.

In \cite{ElectricitymarketsandDR}, the DR problem is decomposed in terms of devices and a waterfilling-inspired negotiation mechanism is proposed to reduce electricity generation costs, whereas \cite{EMSwithDR} proposes a method for the energy management of several prosumers in an energy district in the aim of maximizing the energy district's utility and reducing reverse energy flows. In \cite{Zheng_DRusingLyapunov}, a suboptimal distributed algorithm based on an extended Lyapunov optimization technique is used to control the switching states of HVAC units in the aim of reducing the average variation of nonrenewable energy demand while ensuring user comfort. The works in \cite{DistributedAlgorithmforRDR} and \cite{OptimalDecentralizedRDR} aim at flattening the load profile by minimizing the deviation of the total load in a time-slot from the mean total load over the scheduling horizon subject to the minimum cost of individual consumers. The resulting bi-level optimization problem is transformed into its equivalent single-level problem and solved in a distributed fashion. However, the focus in \cite{DistributedAlgorithmforRDR} and \cite{OptimalDecentralizedRDR} is not on pricing strategies but on demand profile reshaping. Moreoever, \cite{RDRwithinterruptibletasks} shows that the nonconvex demand response problem that results from incorporating devices with interruptible tasks has a zero duality gap if the problem is formulated over a continuous-time horizon. It also shows that, in a discrete-time horizon, the duality gap vanishes as the granularity of the discretization is increased. A conventional gradient method is used in \cite{RDRwithinterruptibletasks} to solve the nonconvex discrete-time DR problem. On the other hand, an approximate greedy iterative algorithm is used in \cite{distributedalgorithmforHEMS} to find sub-optimal energy consumption schedules for the users. Additionally, the algorithm in \cite{distributedalgorithmforHEMS} is guided towards convergence by introducing a penalty term that penalizes large changes between successive iterations. The DR model in \cite{DRwithMIC}, also decomposed in terms of devices, is solved in a distributed fashion using the proximal bundle method.

Similar to \cite{ElectricitymarketsandDR,distributedalgorithmforHEMS} and \cite{DRwithMIC}, the DR problems in \cite{Yaagoubi_userawareGT} and \cite{Tushar_distriburtedRTmechanism} are decomposed in terms of devices but use concepts from game theory to solve the problem. In more detail, a game theoretic approach based on a modified regret matching procedure is proposed in \cite{Yaagoubi_userawareGT} to solve the problem to within $4\%$ of the optimum, whereas \cite{Tushar_distriburtedRTmechanism} formulates the problem as a noncooperative game and uses mechanism design to distributedly solve the problem to a near-optimal Nash equilibrium.

Furthermore, the work in \cite{onlineauctionforDR} proposes a novel auction format, inspired by the clock-proxy auction in \cite{clockproxy}, for the on-line scheduling of large numbers of households and small- and medium-sized businesses, and shows how the mechanism improves the efficiency of on-line energy use scheduling. 

In contrast to \cite{ElectricitymarketsandDR,distributedalgorithmforHEMS,RDRwithinterruptibletasks,DRwithMIC,Yaagoubi_userawareGT} and \cite{Tushar_distriburtedRTmechanism}, the DR problem in this work and in our previous work \cite{distributedalgorithmMIP} is decomposed in terms of households. Doing so, allows for a more expressive household model, which can incorporate the intricate couplings between storage devices, appliances and distributed energy resources. In contrast to \cite{distributedalgorithmMIP}, the method in this paper accounts for user satisfaction and comfort and engages an algorithm that terminates in substantially fewer iterations.

Against this background, this paper proposes a fast distributed gradient algorithm applied to the double smoothed dual function of the adopted DR problem and shows how to recover a near-optimal primal solution. In more detail, this paper advances the state of the art in the following ways:
\begin{itemize}
\item The nonconvex DR problem in this work is decomposed in terms of households, which facilitates incorporating the intricate couplings between storage devices, appliances and distributed energy resources.
\item The proposed distributed gradient algorithm is applied to a double smoothed dual function and is designed to terminate in $60$ iterations, which can be ideal for an on-line version of this problem.
\item Numerical simulations show that, with minimal parameter tuning, the proposed algorithm exhibits a similar convergence behavior throughout all the studied systems and converges to near-optimal solutions, which corroborates its scalability.
\end{itemize}
The paper also provides a deeper insight into the geometry of the dual function of the DR problem and shows that this dual function is nonsmooth. Consequently, the paper demonstrates that a conventional gradient method fails to solve this problem even if the integrality constraints are relaxed and the problem is convex.
% It might seem intuitive that having more continuous variables will result in a tighter optimality gap but in reality, as shown above, even the convex problem dual is nonsmooth and this assumption does not hold for this particular problem.
Taken together, these advances show that the proposed algorithm represents a feasible method for implementing large-scale demand response.

The paper progresses with notation and pertinent concepts from convex optimization in Section~\ref{sec:Preliminaries}, followed by a description of the DR model in Section~\ref{sec:DRmodel}. Sections~\ref{sec:doublesmoothing} describes the double smoothing technique and its properties and Section~\ref{sec:algorithm} presents the proposed fast gradient method. Numerical results are presented in Section~\ref{sec:evaluation} and Section~\ref{sec:conclusion} concludes the paper.

\section{Preliminaries}\label{sec:Preliminaries}
All vectors are column vectors unless otherwise specified, and $\boldsymbol{0}$ is an all-zeros vector of length depending on the context. The inner product of two vectors $\boldsymbol{x}$, $\boldsymbol{y} \in \reals^n$ is delineated by $\left\langle \boldsymbol{x}, \boldsymbol{y} \right\rangle=\boldsymbol{x}' \boldsymbol{y}$, where $\boldsymbol{x}'$ is the transpose of $\boldsymbol{x}$. The Euclidean norm of a vector $\boldsymbol{x} \in \reals^n$ is denoted by $\left\|\boldsymbol{x}\right\|=\sqrt{\left\langle \boldsymbol{x}, \boldsymbol{x} \right\rangle}$ and the nonnegative orthant in $\reals^n$ is denoted by $\reals^n_{+}$. The spectral norm of a matrix $A \in \reals^{n\times m}$ is defined by $\left\|A\right\|=\sqrt{\lambda_{\mathrm{max}}\left(A'A\right)}$, where $\lambda_{\mathrm{max}}\left(A'A\right)$ is the maximum eigenvalue of $A'A$.
 
In smooth convex optimization, $\mathcal{F}^{1,1}_{L}\left(\reals^{n}\right)$ is the class of continuously differentiable convex functions $f:\reals^{n} \mapsto \reals$ with \emph{Lipschitz-continuous} gradient \cite{intoncvxopt}, that is:
\begin{align*}
                \left\|\nabla f\left(\boldsymbol{x}\right)-\nabla f\left(\boldsymbol{y}\right)\right\|\leq L \left\|\boldsymbol{x}-\boldsymbol{y}\right\|, \text{~for all~} \boldsymbol{x},\boldsymbol{y} \in \reals^{n},
\end{align*}
for some constant $L>0$. A continuously differentiable function $f\left(\boldsymbol{x}\right)$ is called \emph{strongly convex} on $\reals^{n}$ (i.e.~$f \in \mathcal{S}^{1}_{\kappa}\left(\reals^{n}\right)$) if there exists a constant $\kappa>0$ such that for any $\boldsymbol{x}$, $\boldsymbol{y} \in \reals^{n}$,
\begin{align*}
  f\left(\boldsymbol{y}\right) \geq f\left(\boldsymbol{x}\right) + \left\langle \nabla f\left(\boldsymbol{x}\right), \boldsymbol{y}-\boldsymbol{x}\right\rangle + \frac{\kappa}{2} \left\|\boldsymbol{y}-\boldsymbol{x}\right\|^2.
\end{align*}

We are particularly interested in functions that belong to the class $\mathcal{S}^{1,1}_{\kappa,L}\left(\reals^{n}\right)$, which is the class of functions $f \in \mathcal{F}^{1,1}_{L}\left(\reals^{n}\right)$ that are strongly convex with parameter $\kappa > 0$.

\section{DR model and problem description}\label{sec:DRmodel}
\begin{figure}[t]
\centering{
 \includegraphics[width=90mm] {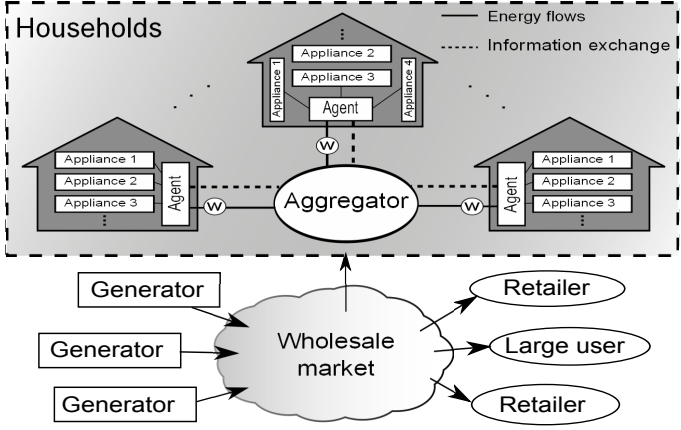}
}
\caption{Aggregator and agents detailed interaction model and the broader energy market (the dashed line is the scope of this paper).}
\label{InteractionModel}
\end{figure}
The adopted DR topology, illustrated in Figure~\ref{InteractionModel}, is composed of one aggregator, which coordinates the schedules of the participating households' loads, interacting with $I$ household agents over a decision horizon $\mathcal{T}:=\{\tau,\tau+\Delta \tau,\ldots,\tau+T-\Delta \tau\}$ (typically one day) consisting of $T$ time-slots. Specifically, the DR model comprises a set of agents $\mathcal{I}:=\{0,1,2, \ldots ,I\}$, where $0$ is the aggregator and each $i \neq 0$ is a household agent. 
 
\subsection{Household agent model}\label{sec:agentmodel}
For each agent $i \neq 0$, let $x^{t}_{i,a}$ be the energy consumption variable of device $a\in \mathcal{A}_{i}:=\{1,\ldots,A_{i}\}$ during time-slot $t$, where $\mathcal{A}_{i}$ is the set of all devices of agent $i$. Each device $a\in \mathcal{A}_{i}$ is associated with a user-defined preferred scheduling interval \mbox{$\tau _{i,a}:=\left\{ \tau^{\mathrm{start}}_{i,a},\ldots,\tau^\mathrm{end}_{i,a}\right\}$}, where $\tau^\mathrm{start}_{i,a}$ and $\tau^\mathrm{end}_{i,a}$ are the start and end times of the desired scheduling interval (e.g.~washing machine desired to be `on' somewhere between 5pm and 9pm or an EV desired to be charged between 11pm and 7am). Devices can be classified into seven types. To this end, let $\phi \in \left\{1,\ldots,7\right\}$ denote the type of agent $i$'s device $a$ and $\mathcal{A}_{i,\phi} \subseteq \mathcal{A}_{i}$ be the set of agent $i$'s type $\phi$ devices. Additionally, let $l \in \{1,\ldots,L\}$ be the operating mode of agent $i$'s device $a \in \left\{\mathcal{A}_{i,1} \cup \mathcal{A}_{i,2} \cup \mathcal{A}_{i,3}\right\}$ and \mbox{$\boldsymbol{p_{i,a}}=[p^{1}_{i,a},\ldots,p^{L}_{i,a}]$} be the associated vector of power levels. Consequently, the energy consumed during $\Delta \tau$ would be $e^{l}_{i,a}= p^{l}_{i,a} \Delta \tau$.

A set of Type 1 includes must-run devices that must always be `on', such as refrigerators. These devices constitute the base load of a household and their feasible set $X_{i,a \in \mathcal{A}_{i,1}} \in \reals^T$ is defined by 
\begin{align}
	x_{i,a}^{t}=e^{1}_{i,a}, \ \ & a \in \mathcal{A}_{i,1}, \ t \in  \mathcal{T}. \label{eq:type1}
\end{align}

A set of Type 2 includes inflexible devices that can operate at discrete power levels, such as electric ovens, lighting and TVs with DVD players or game consoles. Devices of Type 2 do not have a total energy requirement over the scheduling horizon, but they have an adjustable energy level that depends on the dissatisfaction of the user. The feasible scheduling set $X_{i,a \in \mathcal{A}_{i,2}}$ of Type 2 devices is defined by
\begin{align}
	x^{t}_{i,a}-u^{1,t}_{i,a}e^{1}_{i,a}- \cdots -u^{L,t}_{i,a}e^{L}_{i,a}=0, \ \ & a \in \mathcal{A}_{i,2}, \ t \in  \mathcal{T}, \label{eq:type2a} \\
	u^{1,t}_{i,a}+\cdots+u^{L,t}_{i,a}=u^{t}_{i,a}, \ \ \ & a \in \mathcal{A}_{i,2}, \ t \in  \mathcal{T}. \label{eq:type2b}
\end{align}
Constraint \eqref{eq:type2b} restricts only one binary variable $u^{l,t}_{i,a}$ to take a value of `1' during time-slot $t$. Type 2 devices are associated with a function that reflects agent $i$'s tradeoff between cost minimization and satisfaction maximization. This function is defined by
\begin{align}
	D_{i,a}^{t}\left(u^{t}_{i,a}\right)=\gamma^{0}_{i,a} (1-u^{t}_{i,a})+\gamma^{1}_{i,a} u^{1,t}_{i,a}+\cdots+\gamma^{L}_{i,a} u^{L,t}_{i,a}, \nonumber \\
  a \in \mathcal{A}_{i,2}, \ t \in  \tau _{i,a},
\end{align}
%\begin{equation}
%D_{i,a}^{t}\left(u^{t}_{i,a}\right)=
 %\left\{\begin{aligned}
        %& \gamma^{0}_{i,a} (1-u^{t}_{i,a}), &&\text{if } t >  \tau^{\mathrm{end},\pi}_{i,a}, \\
        %& \gamma^{1}_{i,a} u^{1,t}_{i,a}, &&\text{if } t >  \tau^{\mathrm{end},\pi}_{i,a}, \\
				%& \ \ \vdots\\
				%& \gamma^{L}_{i,a} u^{L,t}_{i,a}, &&\text{if } t >  \tau^{\mathrm{end},\pi}_{i,a}, \\
       %\end{aligned}
 %\right.
 %\quad , a \in \mathcal{A}_{i,2}, \ t \in  \tau _{i,a},
%\end{equation}
where $\gamma^{0}_{i,a}$, $\gamma^{1}_{i,a}$, $\ldots$ , $\gamma^{L}_{i,a}$ are nonnegative parameters that reflect agent $i$'s preference for each operating mode $l$. For instance, if a user prefers the highest operating mode over the others, these parameters can be set as $\gamma^{L}_{i,a}=0$ and $\gamma^{L-1}_{i,a} \leq \cdots \leq \gamma^{1}_{i,a} \leq \gamma^{0}_{i,a}$.

A Type 3 set contains flexible and non-interruptible devices whose operation can be delayed or advanced but cannot be interrupted before they have completed their task. Devices of Type 3 have a specific total energy requirement per scheduling horizon. A Type 3 set includes appliances such as dishwashers, washing machines and dryers that can operate at discrete power levels similar to Type 2 devices. In more detail, the feasible scheduling set $X_{i,a \in \mathcal{A}_{i,3}}$ of Type 3 devices is defined by
\begin{align}
	x^{t}_{i,a}-u^{1,t}_{i,a}e^{1}_{i,a}- \cdots -u^{L,t}_{i,a}e^{L}_{i,a}=0, \ \ a \in \mathcal{A}_{i,3}, & \ t \in  \mathcal{T}, \label{eq:type3a} \\
	u^{1,t}_{i,a}+\cdots+u^{L,t}_{i,a}=u^{t}_{i,a}, \ \ a \in \mathcal{A}_{i,3}, & \ t \in  \mathcal{T}, \label{eq:type3b} \\
	\sum_{ t\in \mathcal{T}}x^{t}_{i,a} \geq E_{i,a}, \ \ a \in \mathcal{A}_{i,3}, & \label{eq:appenergy} \\
	v^{t}_{i,a} \geq u^{t}_{i,a}-u^{t-\Delta \tau}_{i,a},\text{\footnotemark} \ a \in \mathcal{A}_{i,3}, & \ t \in  \mathcal{T}, \label{eq:trivialinequality}
\end{align}
\footnotetext{At $t=\tau$, $t-\Delta \tau$ is equal to $\tau-\Delta \tau$, which is equal to $\tau+T-\Delta \tau$ of the previous scheduling horizon.}
 and for all $t\in \{\tau+\pi_{i,a}-\Delta \tau,\ldots,\tau+T-\Delta \tau\}$,
\begin{align}
	\ \sum^{t}_{q=t-\pi_{i,a}+\Delta \tau}v^{q}_{i,a} \leq u^{t}_{i,a}, \ a \in \mathcal{A}_{i,3}. \label{eq:minontime}
\end{align}
The startup binary variable $v^{t}_{i,a}$ is only equal to `1' when device $a$ is turned on during time-slot $t$. The minimum `on' time constraint \eqref{eq:minontime} states that if device $a$ is turned on during time-slot $t$ (i.e. $v^{t}_{i,a}=1$), then this device should remain `on' for at least $\pi_{i,a}$ time-slots. This formulation is not a `hold-time' formulation as the device can still be `on', even after the minimum `on' time has elapsed, in order to fulfill its total energy requirement $E_{i,a}$. Constraints \eqref{eq:trivialinequality} and \eqref{eq:minontime} are inequalities that describe facets of the convex hull of the projection on the space of both $u$ and $v$ \cite{MOTconstraints}. This formulation is a tight polyhedral representation of the convex hull of the disjoint set $X_{i,a \in \mathcal{A}_{i,3}}$. More interestingly, the variable $v$ can be modeled as continuous. Specifically, because $u$ is binary, constraints \eqref{eq:trivialinequality} and \eqref{eq:minontime} ensure that the $v$ variables are binary even if they are modeled as continuous \cite{UCandTS}.

Unlike Type 2 devices, a user only cares that a Type 3 device finishes its task within the preferred scheduling interval $\tau_{i,a}$. Therefore, for Type 3 devices, the dissatisfaction function would be
\begin{equation*}
  D_{i,a}^{t}\left(u^{t}_{i,a}\right)=
	 \left\{\begin{aligned}
        & 0, &&\text{if } t \in  \tau^{\pi}_{i,a}, \ \ \\
        & \overline{\gamma}_{i,a} (t-\tau^{\mathrm{end},\pi}_{i,a})u^{t}_{i,a}, &&\text{if } t >  \tau^{\mathrm{end},\pi}_{i,a}, \\
				& \underline{\gamma}_{i,a}(\tau^{\mathrm{start}}_{i,a}-t)u^{t}_{i,a}, &&\text{if } t <  \tau^{\mathrm{start}}_{i,a},\\
       \end{aligned}
 \right.
\end{equation*}    
\begin{equation}
  \qquad \qquad \qquad \qquad \qquad \qquad \qquad \quad \ a \in \mathcal{A}_{i,3}, \ t \in  \mathcal{T},
\end{equation}
where $\tau^{\pi}_{i,a}=\left\{\tau^{\mathrm{start}}_{i,a},\ldots,\tau^{\mathrm{end},\pi}_{i,a}\right\}$ and $\tau^{\mathrm{end},\pi}_{i,a}=\tau^{\mathrm{end}}_{i,a}+\pi_{i,a}-\Delta \tau$.\footnote{If $\overline{\gamma}_{i,a}=\underline{\gamma}_{i,a}$, function $D_{i,a}^{t}\left(u^{t}_{i,a}\right)$ would be symmetrical around $\tau^{\pi}_{i,a}$.}
Parameters $\overline{\gamma}_{i,a}>0$ and $\underline{\gamma}_{i,a}>0$ determine how quickly the user gets dissatisfied when the scheduled operation of Type 3 device $a$ is delayed by $t-\tau^{\mathrm{end},\pi}_{i,a}$ time-slots away from $\tau^{\mathrm{end}}_{i,a}$ or advanced $\tau^{\mathrm{start}}_{i,a}-t$ time-slots ahead of $\tau^{\mathrm{start}}_{i,a}$, respectively.

A Type 4 set contains flexible and interruptible storage devices with a continuous power level within a certain range $\tau_{i,a}$, like EVs. Their feasible scheduling set $X_{i,a \in \mathcal{A}_{i,4}}$ is defined by
\begin{align}
	u^{\mathrm{ch},t}_{i,a}\left(P^\mathrm{ch,min}_{i,a} \Delta \tau\right) \leq & \ x^{\mathrm{ch},t}_{i,a}\leq u^{\mathrm{ch},t}_{i,a}\left(P^\mathrm{ch,max}_{i,a} \Delta \tau\right), \nonumber \\
	& \ \ \ \ \qquad \qquad a \in \mathcal{A}_{i,4}, \ t \in \tau_{i,a}, \label{eq:type4ch} \\
	u^{\mathrm{dis},t}_{i,a}\left(P^\mathrm{dis,min}_{i,a} \Delta \tau\right) \leq & \ x^{\mathrm{dis},t}_{i,a}\leq u^{\mathrm{dis},t}_{i,a}\left(P^\mathrm{dis,max}_{i,a} \Delta \tau\right), \nonumber \\
	& \ \ \ \ \qquad \qquad a \in \mathcal{A}_{i,4}, \ t \in \tau_{i,a}, \label{eq:type4dis} \\
	u^{\mathrm{dis},t}_{i,a}+u^{\mathrm{ch},t}_{i,a}=& \ u^{t}_{i,a}, \ \ \qquad \ a \in \mathcal{A}_{i,4}, \ t \in \tau_{i,a}, \label{eq:type4coupling} \\
	x^{\mathrm{ch},t}_{i,a}-x^{\mathrm{dis},t}_{i,a}=& \ x^{t}_{i,a},\text{\footnotemark} \ \qquad \ a \in \mathcal{A}_{i,4}, \ t \in \mathcal{T}, \label{eq:definingconst4} \\
	x^{\mathrm{SoC},t}_{i,a}=& \ x^{\mathrm{SoC},t-\Delta \tau}_{i,a}+\eta^{\mathrm{ch}}_{i,a} x^{\mathrm{ch},t}_{i,a}-\frac{x^{\mathrm{dis},t}_{i,a}}{\eta^{\mathrm{dis}}_{i,a}}, \nonumber \\
	& \ \ \ \ \qquad \qquad a \in \mathcal{A}_{i,4}, \ t \in \tau_{i,a}, \label{eq:type4soc} \\
	x^{\mathrm{SoC},\tau^{\mathrm{start}}_{i,a}-\Delta \tau}_{i,a}=& \ e^{\mathrm{SoC,ini}}_{i,a}, \ \ \ \ a \in \mathcal{A}_{i,4},\\
	x^{\mathrm{SoC},\tau^{\mathrm{end}}_{i,a}}_{i,a}=& \ e^{\mathrm{SoC,final}}_{i,a}, \ \ a \in \mathcal{A}_{i,4},\\
	e^{\mathrm{SoC,min}}_{i,a} \leq x^{\mathrm{SoC},t}_{i,a}\leq& \ e^{\mathrm{SoC,max}}_{i,a}, \ \ a \in \mathcal{A}_{i,4}, \ t \in \tau_{i,a}. 
	\end{align}        
%Note that for an EV, $u^{t}_{i,a}=0$ for any $t \notin \tau_{i,a}$.
\footnotetext{\label{note1} Constraints \eqref{eq:definingconst4} and \eqref{eq:definingconst5} are \emph{defining constraints} that only provide a definition to the variables on their left-hand side. They are therefore eliminated during presolve.}
A Type 5 set contains flexible and interruptible storage devices with a continuous power level over $\mathcal{T}$, like batteries. Their feasible scheduling set $X_{i,a \in \mathcal{A}_{i,5}}$ is defined by
\begin{align}
	u^{\mathrm{ch},t}_{i,a}\left(P^\mathrm{ch,min}_{i,a} \Delta \tau\right) \leq & \ x^{\mathrm{ch},t}_{i,a}\leq u^{\mathrm{ch},t}_{i,a}\left(P^\mathrm{ch,max}_{i,a} \Delta \tau\right), \nonumber \\
	& \ \ \ \ \qquad \qquad a \in \mathcal{A}_{i,5}, \ t \in \mathcal{T}, \label{eq:type5ch} \\
	u^{\mathrm{dis},t}_{i,a}\left(P^\mathrm{dis,min}_{i,a} \Delta \tau\right) \leq & \ x^{\mathrm{dis},t}_{i,a}\leq u^{\mathrm{dis},t}_{i,a}\left(P^\mathrm{dis,max}_{i,a} \Delta \tau\right), \nonumber \\
	& \ \ \ \ \qquad \qquad a \in \mathcal{A}_{i,5}, \ t \in \mathcal{T}, \label{eq:type5dis} \\
	u^{\mathrm{dis},t}_{i,a}+u^{\mathrm{ch},t}_{i,a}=& \ u^{t}_{i,a}, \ \ \qquad \ a \in \mathcal{A}_{i,5}, \ t \in \mathcal{T}, \label{eq:type5coupling} \\
	x^{\mathrm{ch},t}_{i,a}-x^{\mathrm{dis},t}_{i,a}=& \ x^{t}_{i,a},\text{\footnotemark[3]} \ \qquad \ a \in \mathcal{A}_{i,5}, \ t \in \mathcal{T}, \label{eq:definingconst5}
	\end{align}
	\begin{align}
	x^{\mathrm{SoC},t}_{i,a}=& \ x^{\mathrm{SoC},t-\Delta \tau}_{i,a}+\eta^{\mathrm{ch}}_{i,a} x^{\mathrm{ch},t}_{i,a}-\frac{x^{\mathrm{dis},t}_{i,a}}{\eta^{\mathrm{dis}}_{i,a}}, \nonumber \\
	& \ \ \ \ \qquad \qquad a \in \mathcal{A}_{i,5}, \ t \in \mathcal{T}, \label{eq:type5soc} \\
	x^{\mathrm{SoC},\tau^{\mathrm{start}}_{i,a}-\Delta \tau}_{i,a}=& \ e^{\mathrm{SoC,ini}}_{i,a}, \ \ \ \ a \in \mathcal{A}_{i,5},\\
	x^{\mathrm{SoC},\tau^{\mathrm{end}}_{i,a}}_{i,a}\geq & \ e^{\mathrm{SoC,final}}_{i,a}, \ \ a \in \mathcal{A}_{i,5},\\
	e^{\mathrm{SoC,min}}_{i,a} \leq x^{\mathrm{SoC},t}_{i,a}\leq& \ e^{\mathrm{SoC,max}}_{i,a}, \ \ a \in \mathcal{A}_{i,5}, \ t \in \mathcal{T}. 
	\end{align}
% In reality, there's a thin line between some of the types and the distinction can be blurred by the idiosynchracies of the user. Therefore, our contention here is to be as generic as possible and to demonstrate that the proposed method works with any mixture of device types.
A Type 6 set contains thermostatically controlled devices like air conditioners. Their feasible scheduling set $X_{i,a \in \mathcal{A}_{i,6}}$ is defined by 
\begin{align}
	u^{t}_{i,a}\left(P^\mathrm{th,min}_{i,a} \Delta \tau\right) \leq & \ x^{t}_{i,a}\leq u^{t}_{i,a}\left(P^\mathrm{th,max}_{i,a} \Delta \tau\right), \nonumber \\
	& \ \ \ \ \qquad \qquad a \in \mathcal{A}_{i,6}, \ t \in \tau_{i,a}, \label{eq:type6U}
\end{align} 
	and their operation is governed by the (first order) thermal dynamics
\begin{align}
T_{i}^{\mathrm{in},t}=T_{i}^{\mathrm{in},t-\Delta \tau}&+\psi_{i,a}  x_{i,a}^{t}+\zeta_{i,a} \left(T^{\mathrm{out},t-\Delta \tau}-T_{i}^{\mathrm{in},t-\Delta \tau}\right), \nonumber \\
	& \ \ \ \qquad \qquad \qquad a \in \mathcal{A}_{i,6}, \ t \in \tau_{i,a}, \label{eq:thermal}	\\
	T_{i}^{\mathrm{in,min}} & \leq T_{i}^{\mathrm{in},t} \leq T_{i}^{\mathrm{in,max}}, \ \ \ \quad t \in \tau_{i,a},
\end{align}
where $\psi_{i,a}>0$ and $\zeta_{i,a}$ are parameters defined by the geometry of the house (or room), thermal properties of the house (or room) materials, and the thermostatically controlled device characteristics (temperature of air flow, air mass flow rate) \cite{householdthermalmodel}. Moreover, $\psi_{i,a}<0$ if device $a \in \mathcal{A}_{i,6}$ is in cooling mode and $\psi_{i,a}>0$ if device $a \in \mathcal{A}_{i,6}$ is in heating mode.

Additionally, Type 6 devices are associated with a user dissatisfaction function captured by 
\begin{align}
	D^{t}_{i,a}\left(T_{i}^{\mathrm{in},t}\right)=\gamma^{\mathrm{th}}_{i,a} \left(T_{i}^{\mathrm{in},t}-T_{i}^{\mathrm{comf}}\right)^2, \nonumber \\
	\qquad \qquad a \in \mathcal{A}_{i,6}, \ t \in \tau_{i,a}. \label{eq:dissatisfaction6}
\end{align}
The dissatisfaction function in \eqref{eq:dissatisfaction6} is adopted from \cite{NaLi} and aims at reflecting agent $i$'s tolerance to deviations of $T_{i}^{\mathrm{in},t}$ from $T_{i}^{\mathrm{comf}}$.

Given the above, the electric energy demand of agent $i \neq 0$ during time-slot $t$ is denoted by $x^{t}_{i} \in X^{t}_{i}$, where $X^{t}_{i}$ is defined by
\begin{align}
  x^{t}_{i}=\sum_{ a\in \mathcal{A} _{i}}x^{t}_{i,a}-P_{i}^{\mathrm{PV},t} \Delta \tau, \label{eq:agenticoupling} \\
	0 \leq x^{t}_{i} \leq P^{\mathrm{max}}_{i}\Delta \tau.  \label{eq:maxenergy}
\end{align}
Constraint \eqref{eq:maxenergy} restricts agent $i$'s total energy consumption during time-slot $t$ to a maximum threshold of $E^{\mathrm{max}}_{i} =P^{\mathrm{max}}_{i}\Delta \tau$. This constraint can be thought of as a way to ensure that the power consumption during time-slot $t$ does not exceed the rated capacity of the household main circuit breaker's overload protection. In fact, one of the superiorities of the method in this paper is its ability to handle the coupling constraints \eqref{eq:agenticoupling} and \eqref{eq:maxenergy}, which can only be incorporated in a distributed model that decomposes the problem in terms of households. 

Finally, the demand profile of agent $i \neq 0$ is denoted by \mbox{$\boldsymbol{x_i}=\left[x_{i}^{\tau}, \ldots , x_{i}^{\tau+T-\Delta \tau} \right]\in X_{i}$},
where $X_{i}=\left(\prod_{a \in \mathcal{A} _{i}} X_{i,a}\right) \times \left(\prod_{t \in \mathcal{T}} X_{i}^{t}\right)$. Because of the presence of binary variables (enforced by integrality constraints), the feasible sets $X_{i}$ are disjoint and therefore nonconvex.

\subsection{Aggregator model}\label{sec:aggregatormodel}
The aggregator purchases energy in a pooled wholesale market, and as such, faces a set of cost functions $C^{t}:\reals_{+} \mapsto \reals_{+}$. In this expression, $C^{t}\left(x^{t}_{g}\right)$ is the cost of drawing $x^{t}_{g}$ units of energy from the grid during time-slot $t$. Due to physical system limits, the power drawn from the grid is bounded above by $G^{\mathrm{max}}$, which represents the maximum power that can be drawn from the grid, and therefore $x^{t}_{g} \in X_{g}:=[0,G^{\mathrm{max}} \Delta \tau]$. Under the assumption that open-cycle gas turbines are the marginal energy producers\footnote{In reality the wholesale prices are also affected by congestions on the transmission network but in this paper this congestion component of the wholesale electricity prices is neglected.}, the cost faced by an aggregator buying energy in the wholesale market during time-slot $t$ can be approximated by the convex quadratic function,
\begin{align}\label{eq:costfunc}
	C^{t}\left(x_{g}^{t}\right)=c2^{t} \left(x_{g}^{t}\right)^{2}+c1^{t}x_{g}^{t}+c0^{t},
\end{align}
where $c0^{t}$, $c1^{t}$ and $c2^{t}$ are time-varying parameters that reflect the fluctuating wholesale prices.\footnote{The cost function in \eqref{eq:costfunc} is not tailored to a specific market but instead is a general approximation of efficient markets.} Throughout the rest of the paper, parameters $c1^{t}$ and $c0^{t}$ are set to 0 for instantiation purposes.\footnote{Strictly positive values of $c1^{t}$ and $c0^{t}$ do not affect the derivations in this paper.}

\subsection{Demand aggregation problem}\label{sec:aggregationproblem}
The feasible scheduling sets $X_{i\neq 0}$ are private information held individually by each household. If the aggregator is able to access this information for all $i \neq 0$, then it can (centrally) minimize the total energy cost per scheduling horizon $\mathcal{T}$, and thereby efficiently allocate electric energy to these households, by solving the following problem:
\begin{subequations}\label{eq:mincost}
\begin{align}
& \underset {\substack{
	\boldsymbol{x_{i}} \in X_{i},x_{g}^{t} \in X_{g},\\
	u^{t}_{i,a},T^{\mathrm{in},t}_{i}}} {\mbox{minimize}} \qquad
   \sum_{ t\in \mathcal{T}} \Bigg(C^{t}\left(x^{t}_{g}\right)+\Bigg. \nonumber \\
& \left. \sum_{i \in \mathcal{I} \setminus 0}\left(\sum_{a \in \mathcal{A} _{i,2} \cup \mathcal{A} _{i,3}}  D_{i,a}^{t}\left(u^{t}_{i,a}\right)+\sum_{a \in \mathcal{A} _{i,6}}D_{i,a}^{t}\left(T^{\mathrm{in},t}_{i}\right)\right)\right), \\
& \text{subject to} 
  \qquad  \sum_{ i\in \mathcal{I} \setminus 0} x^{t}_{i}=x^{t}_{0}, \qquad t \in \mathcal{T}, \label{eq:coupling}
\end{align}
\end{subequations}
where $x_{0}^{t}=x^{t}_{g}$ and $x^{t}_{0} \in X_{0} \in \reals^{T}_{+}$ is the total demand during time-slot $t$. 

Letting 
$\boldsymbol{x} = \left\{
  \left\{\boldsymbol{x_i}\right\}_{i\in \mathcal{I}},
  \left\{x_g^t\right\}_{t\in \mathcal{T}}\right\}$, and with a slight abuse of notation, problem \eqref{eq:mincost} can also be written as:
\begin{align}
\mathcal{P}^{*}
  =\inf_{\boldsymbol{x} \in X} \left\{
  C\left(\boldsymbol{x}\right)+D\left(\boldsymbol{x}\right) \::\:
  A_{c}\boldsymbol{x}=\boldsymbol{0}\right\},
\end{align}
where $X =\left(\prod_{ i\in \mathcal{I}} X_{i}\right) \times X_{g}$ and $A_{c} \in \reals^{T\times[\left(I+1\right)\times T+T]}$ is the coupling constraint matrix concatenating constraints \eqref{eq:coupling}.

Problem \eqref{eq:mincost} is a mixed-integer quadratic program (MIQP) that belongs to the class of NP-hard problems that are notorious for tending to be intractable (if solved centrally for optimality) when they grow in size. In addition, sending the households' private information to the aggregator requires a large communication overhead in a setting with a large number of household agents, even before privacy issues are considered.

However, relaxing the coupling constraints \eqref{eq:coupling} through the Lagrangian relaxation method bestows a separable structure on problem \eqref{eq:mincost}. The problem can then be decomposed into $I+1$ independent subproblems that can be solved in parallel. 

In more detail, the partial Lagrangian of \eqref{eq:mincost} is given by:
\begin{align*}
  \mathcal{L}\left(\boldsymbol{x},\boldsymbol{\lambda} \right) 
  & = \sum_{ t\in \mathcal{T}} \left(C^{t}\left(x^{t}_{g}\right)
     + \lambda^{t} \left(\sum_{ i\in \mathcal{I} \setminus 0} x^{t}_{i} - x^{t}_{0}\right)+\right. \nonumber \\
	& \hspace{-1.5cm} \left. \sum_{i \in \mathcal{I} \setminus 0}\left(\sum_{a \in \mathcal{A} _{i,2} \cup \mathcal{A} _{i,3}}  D_{i,a}^{t}\left(u^{t}_{i,a}\right)+\sum_{a \in \mathcal{A} _{i,6}}D_{i,a}^{t}\left(T^{\mathrm{in},t}_{i}\right)\right)\right) \\
  & = \sum_{ t\in \mathcal{T}} \left(C^{t}\left(x^{t}_{g}\right) - \lambda^{t}x^{t}_{0} \right) 
     +\sum_{ i\in \mathcal{I} \setminus 0} \Bigg(\sum_{ t\in \mathcal{T}} \left( \lambda^{t} x^{t}_{i} +\Bigg. \right.\\
		& \left. \left. \sum_{a \in \mathcal{A} _{i,2} \cup \mathcal{A} _{i,3}}  D_{i,a}^{t}\left(u^{t}_{i,a}\right)+\sum_{a \in \mathcal{A} _{i,6}}D_{i,a}^{t}\left(T^{\mathrm{in},t}_{i}\right)\right)\right),
\end{align*}
where $\boldsymbol{\lambda}=\left[\lambda^{\tau}, \ldots , \lambda^{\tau+T-\Delta \tau} \right]$ is the vector of Lagrange multipliers. Accordingly, the Lagrange dual function is
\begin{align}\label{eq:LagrangeDual}
\mathcal{D}\left(\boldsymbol{\lambda}\right) 
  = & \underset {\substack{\boldsymbol{x} \in X}} {\mbox{inf}} 
    \quad  \mathcal{L}\left(\boldsymbol{x},\boldsymbol{\lambda} \right). 
\end{align}

Due to the block angular structure of the primal problem, elements of the Lagrange dual \eqref{eq:LagrangeDual} can be separated as follows:
\begin{align}
\mathcal{D}\left(\boldsymbol{\lambda}\right) 
   = \mathcal{D}_{0}\left(\boldsymbol{\lambda}\right)
      +\sum_{ i\in \mathcal{I}\setminus 0} \mathcal{D}_{i}\left(\boldsymbol{\lambda}\right),
\end{align}
where the aggregator solves
\begin{align} \label{eq:subproblem0}
\mathcal{D}_{0}\left(\boldsymbol{\lambda}\right)
  = & \underset {\substack{\boldsymbol{x_{0}} \in X_{0},\\x_{g}^{t} \in X_{g}}} {\mbox{inf}} 
    \quad \sum_{ t\in \mathcal{T}} \left(C^{t}\left(x^{t}_{g}\right) - \lambda^{t}x^{t}_{0} \right), 
\end{align}
while the household agents solve
\begin{align} \label{eq:subproblems}
\mathcal{D}_{i}\left(\boldsymbol{\lambda}\right)
  =& \underset {\substack{\boldsymbol{x_{i}} \in X_{i},\\u^{t}_{i,a},T^{\mathrm{in},t}_{i}}} {\mbox{inf}} \sum_{ t\in \mathcal{T}} \left(\lambda^{t} x^{t}_{i}+\sum_{a \in \mathcal{A} _{i,2} \cup \mathcal{A} _{i,3}}  D_{i,a}^{t}\left(u^{t}_{i,a}\right)\right. \nonumber \\
	& \left. +\sum_{a \in \mathcal{A} _{i,6}}D_{i,a}^{t}\left(T^{\mathrm{in},t}_{i}\right)\right), \quad i \in \mathcal{I} \setminus 0. 
\end{align}
Finally, the dual problem is given by
\begin{align}\label{eq:dualproblem}
  \max_{\boldsymbol{\lambda} \succeq \boldsymbol{0}} \quad \mathcal{D}\left(\boldsymbol{\lambda}\right).
\end{align}

\begin{figure}[t]
\centering{
 \psfrag{D}[b][t]{\footnotesize $\mathcal{D} \left(\lambda^{1},\lambda^{2}\right)$ \normalsize}
 \psfrag{1}[t][b]{\footnotesize $\lambda^{1}$ \normalsize}
 \psfrag{2}[t][b]{\footnotesize $\lambda^{2}$ \normalsize}
 \includegraphics[width=90mm] {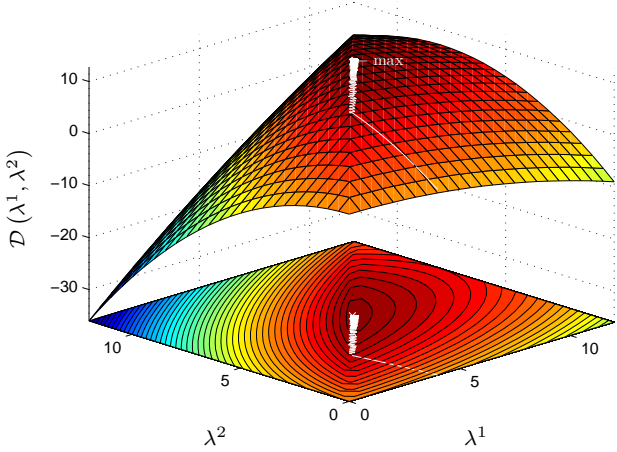}
}
\caption{Dual function $\mathcal{D} \left(\lambda^{1},\lambda^{2}\right)$ (and its contour plot) of a small DR problem with two time slots. The white line delineates the evolution of the dual iterates.}
\label{dualplot}
\end{figure}

However, in this DR scenario, the concave dual function $\mathcal{D}\left(\boldsymbol{\lambda}\right)$ is typically nondifferentiable. Indeed, using Danskin's theorem \cite{Danskin,onatheoremofDanskin,nonlinearpogramming}, the subdifferentials of $\mathcal{D}\left(\boldsymbol{\lambda}\right)$ are
\begin{align*}
  \partial \mathcal{D}\left(\boldsymbol{\lambda}\right)
  =	\left\{A_{c}\boldsymbol{x}:\mathcal{D}\left(\boldsymbol{\lambda}\right) , \boldsymbol{x} \in X \right\}.
\end{align*}
Specifically, as the subproblems in \eqref{eq:subproblem0} and~\eqref{eq:subproblems} can have multiple optimal solutions for a given vector $\boldsymbol{\lambda}$, the subdifferentials $\partial \mathcal{D}\left(\boldsymbol{\lambda}\right)$ may be not be unique and the dual function $\mathcal{D}\left(\boldsymbol{\lambda}\right)$ can be nonsmooth.\footnote{If a function $f\left(\boldsymbol{x}\right)$ is smooth, its subdifferential contains only one point and therefore $\partial f\left(\boldsymbol{x}\right)=\nabla f\left(\boldsymbol{x}\right)$.} Consequently, applying a conventional gradient method \cite{subgradientmethods} to this problem would most likely exhibit very slow convergence. This can be visualized in Figure~\ref{dualplot}, which illustrates the concave but nonsmooth dual function (and its contour plot) of a small problem comprising two households, each with two devices (an EV and an electric oven) scheduled over two time slots. Figure~\ref{dualplot} also showcases the slow convergence of a conventional gradient method as delineated by the white line.
Therefore, in order to accelerate convergence, a double smoothing technique is introduced, which involves regularizing the dual problem in \eqref{eq:dualproblem} to allow applying a fast gradient method \cite{intoncvxopt}.

\section{Double smoothing method}\label{sec:doublesmoothing}
As discussed in Section~\ref{sec:aggregationproblem}, the dual function of the DR problem at hand is typically nonsmooth and not strongly convex. However, these properties can be conferred on the dual function $\mathcal{D}\left(\boldsymbol{\lambda}\right)$ by applying a double smoothing technique.

\subsection{First smoothing}
One way to obtain a smooth approximation of $\mathcal{D}\left(\boldsymbol{\lambda}\right)$ is to modify the subproblems in \eqref{eq:subproblems} to ensure a unique optimal solution for every $\boldsymbol{\lambda}$. The dual function is modified as follows:
\begin{align}
	\mathcal{D}_{\mu}& \left(\boldsymbol{\lambda}\right) =\mathcal{D}_{0}\left(\boldsymbol{\lambda}\right)+\sum_{ i\in \mathcal{I} \setminus 0} \mathcal{D}_{i,\mu} \left(\boldsymbol{\lambda}\right),  \label{eq:smootheddual}
\end{align}
where
\begin{align}
	\mathcal{D}_{i,\mu} & \left(\boldsymbol{\lambda}\right)
  = \underset {\substack{\boldsymbol{x_{i}} \in X_{i},\\u^{t}_{i,a},T^{\mathrm{in},t}_{i}}} {\mbox{inf}} \left(\sum_{ t\in \mathcal{T}} \left(\lambda^{t} x^{t}_{i}+\sum_{a \in \mathcal{A} _{i,2} \cup \mathcal{A} _{i,3}}  D_{i,a}^{t}\left(u^{t}_{i,a}\right)\right. \right. \nonumber \\
	& \hspace{-0.5cm} \left. \left. +\sum_{a \in \mathcal{A} _{i,6}}D_{i,a}^{t}\left(T^{\mathrm{in},t}_{i}\right)\right)+\frac{\mu}{2}\left\|\boldsymbol{x_i}\right\|^2\right), \quad i \in \mathcal{I} \setminus 0. \label{eq:smoothsubproblem}
\end{align}
and $\mu > 0$ is a smoothness parameter \cite{smoothminimization}.
The modified dual function $\mathcal{D}_{\mu}\left(\boldsymbol{\lambda}\right)$ is smooth and its gradient $\nabla \mathcal{D}_{\mu}\left(\boldsymbol{\lambda}\right)=A_{c} \boldsymbol{x}_{\mu,\boldsymbol{\lambda}}$, where $\boldsymbol{x}_{\mu,\boldsymbol{\lambda}}$ delineates the unique optimal solution of problem \eqref{eq:smootheddual}, is Lipschitz-continuous with Lipschitz constant $L_{\mu}=\frac{\left\|A_{c}\right\|^2}{\mu}$. To show the bounds introduced by $\mathcal{D}_{\mu} \left(\boldsymbol{\lambda}\right)$ on $\mathcal{D} \left(\boldsymbol{\lambda}\right)$, let 
\begin{align*}
	D_{X}=\min \left\{ \frac{1}{2} \left\| \left\{\boldsymbol{x_i}\right\}_{i\in \mathcal{I} \setminus 0} \right\|^{2} : \left\{\boldsymbol{x_i}\right\}_{i\in \mathcal{I} \setminus 0} \in \prod_{ i \in \mathcal{I} \setminus 0} X_{i} \right\},
\end{align*}
then $\mathcal{D}_{\mu} \left(\boldsymbol{\lambda}\right) -\mu D_{X} \leq \mathcal{D} \left(\boldsymbol{\lambda}\right) \leq \mathcal{D}_{\mu} \left(\boldsymbol{\lambda}\right)$ for all $\boldsymbol{\lambda} \in \reals_{+}^{T}$ and $-\mathcal{D}_{\mu} \left(\boldsymbol{\lambda}\right) \in \mathcal{F}^{1,1}_{L_{\mu}}$.

The aim of this smoothing is to obtain a Lipschitz-continuous gradient for which efficient smooth optimization methods can be applied. However, despite having a good convergence rate of $\mathcal{D}_{\mu}\left(\boldsymbol{\lambda}^{*}\right)-\mathcal{D}_{\mu}\left(\boldsymbol{\lambda}_{k}\right)$ at iteration $k$ when applying a fast gradient method, the same good rate of convergence does not apply to $\left\|\nabla \mathcal{D}_{\mu}\left(\boldsymbol{\lambda}_{k}\right)\right\|$. Moreover, since the aim is not only to efficiently solve the dual problem but also to recover a feasible solution to the primal, a single smoothing is not enough to achieve this goal \cite{doublesmoothing}. 

\subsection{Second smoothing}
\begin{figure}[t]
\centering{
\psfrag{D}[b][t]{\footnotesize $\mathcal{D}_{\mu,\kappa} \left(\lambda^{1},\lambda^{2}\right)$ \normalsize}
\psfrag{1}[t][b]{\footnotesize $\lambda^{1}$ \normalsize}
\psfrag{2}[t][b]{\footnotesize $\lambda^{2}$ \normalsize}
\includegraphics[width=90mm] {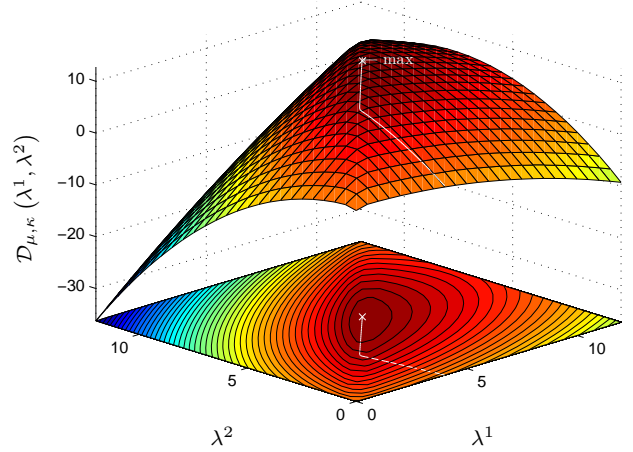}}
\caption{Double smoothed dual function $\mathcal{D}_{\mu,\kappa} \left(\lambda^{1},\lambda^{2}\right)$ (and its contour plot) of the DR example in Section~\ref{sec:aggregationproblem} for $\mu=0.2$ and $\kappa=0.02$. The white line delineates the evolution of the double smoothed dual iterates.}
\label{doublesmootheddualplot}
\end{figure}
This goal can be achieved by applying a second smoothing to the dual function to make it strongly concave. The new dual function is written as
\begin{align}\label{eq:doublesmootheddual}
	\mathcal{D}_{\mu,\kappa} \left(\boldsymbol{\lambda}\right) =\mathcal{D}_{0}\left(\boldsymbol{\lambda}\right)+\sum_{ i\in \mathcal{I} \setminus 0} \mathcal{D}_{i,\mu} \left(\boldsymbol{\lambda}\right)-\frac{\kappa}{2}\left\|\boldsymbol{\lambda}\right\|^{2},
\end{align}
which is strongly concave with parameter $\kappa > 0$, and whose gradient $\nabla \mathcal{D}_{\mu,\kappa}\left(\boldsymbol{\lambda}\right) = A_c \boldsymbol{x}_{\mu,\boldsymbol{\lambda}} - \kappa \boldsymbol{\lambda}$ is Lipschitz-continuous with constant $L_{\mu,\kappa}=\frac{\left\|A_{c}\right\|^2}{\mu}+\kappa=L_{\mu}+\kappa$. Now that $-\mathcal{D}_{\mu,\kappa} \left(\boldsymbol{\lambda}\right) \in \mathcal{S}^{1,1}_{\kappa,L_{\mu,\kappa}}$, applying a fast gradient method ensures the same rate of convergence for $\left\|\nabla \mathcal{D}_{\mu,\kappa}\left(\boldsymbol{\lambda}\right)\right\|$ as for $\mathcal{D}_{\mu,\kappa}\left(\boldsymbol{\lambda}^{*}\right)-\mathcal{D}_{\mu,\kappa}\left(\boldsymbol{\lambda}_{k}\right)$. This property is essential for recovering a near-optimal solution for the primal in fewer iterations compared to just applying a single smoothing \cite{doublesmoothing}. 

The effect of the double smoothing is showcased in Figure~\ref{doublesmootheddualplot} which illustrates the double smoothed dual function $\mathcal{D}_{\mu,\kappa} \left(\lambda_{1},\lambda_{2}\right)$ (and its contour plot) of the DR example in Section~\ref{sec:aggregationproblem} with $\mu=0.2$ and $\kappa=0.02$. Figure~\ref{doublesmootheddualplot} also shows a better performance of the conventional gradient method now applied to the double smoothed dual problem, as delineated by the white line.  

\section{Fast gradient algorithm}\label{sec:algorithm}
The algorithm is divided into two phases. The first phase consists of a fast gradient method applied to the double smoothed dual function in \eqref{eq:doublesmootheddual}. The fast gradient method in the first phase is designed to run for a fixed number of iterations during which both the recovered primal objective value and the norm of the gradient of the dual function are quickly decreased. At the termination of Phase I, the vector of Lagrange multipliers along with the smoothness parameter and the step size that resulted in the minimum recovered primal objective value are selected as a warm start for second phase. In the second phase, the second smoothing is dropped and a penalty term is added to the single smoothed dual function in \eqref{eq:smootheddual}. The penalty term in Phase II penalizes large deviations of household agent $i$'s total load at time-slot $t$ from its value at the previous iteration. Similar to Phase I, Phase II is also designed to run for a fixed number of iterations irrespective of the size of the DR system. 

More specifically, the fast gradient method in Phase I involves two multiplier updates, 
\begin{align}
	\boldsymbol{\lambda}_{k+1}=\hat{\boldsymbol{\lambda}}_{k}+\frac{1}{L_{\mu^k,\kappa^k}^{k}} \nabla \mathcal{D}_{\mu^k,\kappa^k}\left(\hat{\boldsymbol{\lambda}}_{k}\right), \label{eq:lambda1}\\
	\hat{\boldsymbol{\lambda}}_{k+1}=\boldsymbol{\lambda}_{k+1}+\beta^{k} \left(\boldsymbol{\lambda}_{k+1}-\boldsymbol{\lambda}_{k}\right), \label{eq:lambda2}
\end{align}
where 
\begin{align}
	\beta^{k}=\frac{\left(\sqrt{L_{\mu^k,\kappa^k}^{k}}-\sqrt{\kappa^{k}}\right)}{\left(\sqrt{L_{\mu^k,\kappa^k}^{k}}+\sqrt{\kappa^{k}}\right)}.
\end{align}
The parameters of Phase I are set as follows:
\begin{align}
	\mu^{k+1}=\mathrm{e}^{\left(\frac{\log\left(\frac{\mu^{\text{min}}}{\mu^{1}}\right)}{\text{2maxiterI}}\right)} \mu^{k}, \ \kappa^{k+1}= \mathrm{e}^{\left(\frac{\log\left(\frac{\kappa^{\text{min}}}{\kappa^{1}}\right)}{\text{3maxiterI}}\right)} \kappa^{k}, \label{eq:parameters}
\end{align}
where $\text{maxiterI}$ is the maximum number of iterations in Phase I.

In Phase II, the single smoothed dual is modified to incorporate the penalty term as follows:
\begin{align}
	\mathcal{D}_{\mu^{k},\nu^{k}}& \left(\boldsymbol{\lambda}_{k}\right) =\mathcal{D}_{0}\left(\boldsymbol{\lambda}_{k}\right)+\sum_{ i\in \mathcal{I} \setminus 0} \mathcal{D}_{i,\mu^{k},\nu^{k}} \left(\boldsymbol{\lambda}_{k}\right),  \label{eq:smootheddualp}
\end{align}
where
\begin{align}
	\mathcal{D}_{i,\mu^{k},\nu^{k}} & \left(\boldsymbol{\lambda}_{k}\right)
  = \underset {\substack{\boldsymbol{x_{i}} \in X_{i},\\u^{t}_{i,a},T^{\mathrm{in},t}_{i}}} {\mbox{inf}} \left(\sum_{ t\in \mathcal{T}} \Bigg(\lambda_{k}^{t} x^{t}_{i}+ \Bigg. \right. 	\nonumber \\
	& \hspace{-0.5cm} \left. \left. \sum_{a \in \mathcal{A} _{i,2} \cup \mathcal{A} _{i,3}}  D_{i,a}^{t}\left(u^{t}_{i,a}\right) +\sum_{a \in \mathcal{A} _{i,6}}D_{i,a}^{t}\left(T^{\mathrm{in},t}_{i}\right)\right) + \right. \nonumber \\
	& \hspace{-0.3cm} \left.\frac{\mu^{k}}{2}\left\|\boldsymbol{x_i}\right\|^2+\frac{\nu^{k}}{2}\left\|\boldsymbol{x_i}-\boldsymbol{x_i}^{k-1}\right\|^2\right), \ i \in \mathcal{I} \setminus 0. \label{eq:smoothsubproblemp}
\end{align}

The distributed algorithm is described in Algorithm~\ref{algorithm}.
 \begin{algorithm}
 \caption{Distributed algorithm}
 \begin{algorithmic}[1]
	\STATEx \qquad \qquad \qquad \qquad \quad \ \underline{Phase I}
	\scriptsize
 \renewcommand{\algorithmicrequire}{\textbf{Parameters:}}
 \REQUIRE $\boldsymbol{\lambda}_{1} \succeq \boldsymbol{0} $, $\kappa^{\text{min}}>0$, $\kappa^{1}>>\kappa^{\text{min}}$, $\alpha^{1} \in \left[10^{-3},5\times10^{-4}\right]$, $\alpha^{\text{min}} \in \left[8 \times 10^{-4},10^{-6}\right]$, $\text{maxiterI}>10$.
 \STATE \textbf{Initialization:} Aggregator sets $k=1$, $\mu^{1}=\alpha^{1}\left\|A_c\right\|^2$, $\mu^{\text{min}}=\alpha^{\text{min}}\left\|A_c\right\|^2$, $\hat{\boldsymbol{\lambda}}_{1}=\boldsymbol{\lambda}_{1}$, $\nu^{1}=0$ and $J=1$.
  \WHILE {$k \leq \text{maxiterI}$}
  \STATE Aggregator solves $\mathcal{D}_{0} \left(\boldsymbol{\hat{\lambda}}_{k}\right)$ and broadcasts $\boldsymbol{\hat{\lambda}}_{k}$, $\mu^{k}$ and $\nu^{k}$ to the households.
	\STATE Households solve and return $\mathcal{D}_{i,\mu^{k},\nu^{k}} \left(\boldsymbol{\hat{\lambda}}_{k}\right)$ and $\boldsymbol{x}_{\boldsymbol{i}}$ to the aggregator.
	\STATE Aggregator computes $\nabla \mathcal{D}_{\mu^{k},\nu^{k},\kappa^{k}}\left(\hat{\boldsymbol{\lambda}}_{k}\right)$ and $\mathcal{P}_{r}^{k}$ as in \eqref{eq:recoveredprimal}.
  \STATE 	\parbox[t]{\dimexpr\linewidth-\algorithmicindent}{Aggregator computes $L_{\mu^{k},\kappa^{k}}^{k}=\frac{\left\|A_c\right\|^2}{\mu^{k}}+\kappa^{k}$, sets $\nu^{k+1}=0$ and updates $\boldsymbol{\lambda}_{k+1}$, $\hat{\boldsymbol{\lambda}}_{k+1}$, $\mu^{k+1}$ and $\kappa^{k+1}$.\strut}
	\STATE $k \leftarrow k + 1 $.
	\ENDWHILE
 \renewcommand{\algorithmicensure}{\textbf{End of Phase I:}}
 \ENSURE  Aggregator finds the best primal solution $\mathcal{P}_{r}^{J}$ along with $\boldsymbol{\hat{\lambda}}_{J}$, $\mu^{J}$, $L_{\mu^{J},\kappa^{J}}^{J}$ and $\boldsymbol{x}_{\mu^{J},\nu^{J},\boldsymbol{\hat{\lambda}}_{J}}$ such that $J:=\left\{k:\mathcal{P}_{r}^{J}=\min \left\{ \left\{\mathcal{P}_{r}^{k}\right\}_{k \in \left\{1,\ldots,\text{maxiterI}\right\} }  \right\}\right\}$. 
	\STATEx
	\algrule
	\normalsize
	\STATEx \qquad \qquad \qquad \qquad \quad \ \underline{Phase II}
	\scriptsize
	\renewcommand{\algorithmicrequire}{\textbf{Parameters:}}
 \REQUIRE $\text{maxiterII} >1$.
	 \STATE \textbf{Initialization:} Aggregator sets $\delta=\left(L_{\mu^{J},\kappa^{J}}^{J}\right)^{-1}$, $\boldsymbol{\hat{\lambda}}_{k}=\boldsymbol{\hat{\lambda}}_{J}$, $\mu=\rho \mu^{J}$, $\nu=\sigma \mu^{J}$.
	\WHILE {$k \leq \left(\text{maxiterI+maxiterII}\right)$}
  \STATE Aggregator solves $\mathcal{D}_{0} \left(\boldsymbol{\hat{\lambda}}_{k}\right)$ and broadcasts $\boldsymbol{\hat{\lambda}}_{k}$, $\mu$ and $\nu$ to the households.
	\STATE Households solve and return $\mathcal{D}_{i,\mu,\nu} \left(\boldsymbol{\hat{\lambda}}_{k}\right)$ and $\boldsymbol{x}_{\boldsymbol{i}}$ to the aggregator.
	\STATE Aggregator computes $\nabla \mathcal{D}_{\mu,\nu}\left(\hat{\boldsymbol{\lambda}}_{k}\right)$ and $\mathcal{P}_{r}^{k}$ as in \eqref{eq:recoveredprimal}.
  \STATE Aggregator sets $\hat{\boldsymbol{\lambda}}_{k+1}=\hat{\boldsymbol{\lambda}}_{k}+\delta \nabla \mathcal{D}_{\mu,\nu}\left(\hat{\boldsymbol{\lambda}}_{k}\right)$.
	\STATE $k \leftarrow k + 1 $.
	\ENDWHILE
	\renewcommand{\algorithmicensure}{\textbf{End of Phase II:}}
	\ENSURE Aggregator finds the best primal solution $\mathcal{P}_{r}^{S}$ along with $\boldsymbol{\hat{\lambda}}_{S}$ and $\boldsymbol{x}_{\mu,\nu,\boldsymbol{\hat{\lambda}}_{S}}$ such that $S:=\left\{k:\mathcal{P}_{r}^{S}=\min \left\{ \left\{\mathcal{P}_{r}^{k}\right\}_{k \in \left\{1,\ldots,\text{maxiterI+maxiterII}\right\} }  \right\}\right\}$.
\end{algorithmic} 
\label{algorithm}
 \end{algorithm}

In general, a feasible primal solution can only be recovered when both the dual and the norm of its gradient converge, i.e. when $\mathcal{D}_{\mu^{k},\nu^{k},\kappa^{k}}\left(\hat{\boldsymbol{\lambda}}_{k}\right)-\mathcal{D}_{\mu^{k-1},\nu^{k-1},\kappa^{k-1}}\left(\hat{\boldsymbol{\lambda}}_{k-1}\right)\leq \epsilon$ and $\left\|\nabla \mathcal{D}_{\mu^{k},\nu^{k},\kappa^{k}}\left(\hat{\boldsymbol{\lambda}}_{k}\right)\right\|\leq\epsilon$.\footnote{$\epsilon$ is a small positive number in the order of $10^{-4}$.} In addition, recovering a feasible primal solution when the norm of the gradient of the dual is not equal to zero is nontrivial in general. However, in this DR scenario, the aggregator is purchasing electricity for the households only after receiving their demand profiles, computed as a best response to the price signal $\hat{\boldsymbol{\lambda}}_{k}$. Therefore, the aggregator can in practice force the coupling variable $x_{0}^{t}$ to be equal to $\sum_{ i\in \mathcal{I} \setminus 0} x^{t}_{i,\mu^{k},\nu^{k},\boldsymbol{\hat{\lambda}}_{k}}$ at each time-slot $t$ and solve the following problem:
\begin{subequations}\label{eq:recoveredprimal}
\begin{align}
\mathcal{P}_{r}^{k}&= \underset {\substack{
	x_{g}^{t} \in X_{g}}} {\mbox{minimize}} \ \sum_{ t\in \mathcal{T}} C^{t}\left(x^{t}_{g}\right)+\sum_{ i\in \mathcal{I} \setminus 0} \mathcal{D}_{i,\mu^{k},\nu^{k}} \left(\boldsymbol{\hat{\lambda}}_{k}\right) \nonumber \\
	& \hspace{-0.3cm} -\left\langle \boldsymbol{\hat{\lambda}}_{k},\boldsymbol{x_i}^{k}\right\rangle -\frac{\mu^{k}}{2}\left\|\boldsymbol{x_i}^{k}\right\|^2-\frac{\nu^{k}}{2}\left\|\boldsymbol{x_i}^{k}-\boldsymbol{x_i}^{k-1}\right\|^2 \\
& \text{subject to} \ \  \sum_{ i\in \mathcal{I} \setminus 0} x^{t}_{i,\mu^{k},\nu^{k},\boldsymbol{\hat{\lambda}}_{k}}=x^{t}_{g}, \ t \in \mathcal{T}. 
\end{align}
\end{subequations}
This recovered primal solution $\mathcal{P}_{r}^{k}$ is only feasible when $x^{t}_{g} \in X_{g}:=[0,G^{\mathrm{max}} \Delta \tau]$. However, Phase I of the algorithm is designed to quickly decrease the norm of the gradient of the dual function and move away from potential infeasibility. Moreover, in the highly improbable case where this constraint is violated, typically at the start of the algorithm, the aggregator can still track the evolution of these recovered primal iterates by relaxing this constraint. Eventually, at the termination of the algorithm, the aggregator is able to select a feasible recovered primal solution that achieves the minimum value among the recovered primal iterates, as described at the end of Phase II of Algorithm~\ref{algorithm}. In fact, the aggregator does not need $\sum_{ i\in \mathcal{I} \setminus 0} \mathcal{D}_{i,\mu^{k},\nu^{k}} \left(\boldsymbol{\hat{\lambda}}_{k}\right)$ in order to find the point $\boldsymbol{\hat{\lambda}}_{S}$ that gives the minimum recovered primal value $\mathcal{P}_{r}^{S}$ among the recovered primal iterates. The values of $\sum_{ i\in \mathcal{I} \setminus 0} \mathcal{D}_{i,\mu^{k},\nu^{k}} \left(\boldsymbol{\hat{\lambda}}_{k}\right)$ are retrieved here only for comparison purposes.

\section{Numerical evaluation}\label{sec:evaluation}
The simulations are carried out with $T=24$ hours, $\Delta \tau=\SI{1}{\hour}$ and up to $I=2560$ household agents interacting with one aggregator, as in Figure~\ref{InteractionModel}. Each household has up to 10 devices on average, mixed among the types described in Section~\ref{sec:agentmodel}. Appliances' power levels $p^{l}_{i,a}$ are obtained from Ausgrid's device usage guide \cite{ausgrid} for different manufacturers of the same appliance type and different household data. As a result, for Type 1 appliances, $p^{l}_{i,a}$ is selected randomly from $\left[0.08,0.15\right]$. For Type 2 and Type 3 appliances, $p^{l}_{i,a}$ is selected randomly from $\left[0.1,0.275\right]$ and $\left[0.7,4\right]$ respectively. Also, Type 2 and Type 3 appliances can operate in up to 3 operating modes, i.e. $l \in \left\{1,2,3\right\}$. Moreover, each household has 3 Type 3 appliances on average with $\pi_{i,a}$ selected randomly from the set $\left\{2,3\right\}$ and $E_{i,a}\geq \pi_{i,a} \Delta \tau \mathrm{max}\left\{\left\{p^{l}_{i,a}\right\}_{l \in \left\{1,2,3\right\}}\right\}$. The dissatisfaction parameters $\gamma^{l}_{i,a}$ for Type 2 appliances are selected randomly from $\left[0.001,0.15\right]$, whereas for Type 3 $\overline{\gamma}_{i,a}$ is selected randomly from $\left[0.001,0.15\right]$ with $\underline{\gamma}_{i,a}=1.5 \overline{\gamma}_{i,a}$, which makes the dissatisfaction function for Type 3 devices asymmetrical.

For Type 4 (EVs) and Type 5 (batteries) devices, the values for $e_{i,a}^{\mathrm{SoC,max}}$ are drawn randomly from $\left[9,16\right]$ and $\left[8,11\right]$ respectively, whereas $e_{i,a}^{\mathrm{SoC,min}}$ is set to $0.25e_{i,a}^{\mathrm{SoC,max}}$ for both EVs and batteries to avoid deep discharging. The minimum and maximum charging powers $P^\mathrm{ch,min}_{i,a}$ and $P^\mathrm{ch,max}_{i,a}$ for both EVs and batteries are drawn randomly from $\left[0.1,0.6\right]$ and $\left[1.1,3.3\right]$ respectively. Analogously, $P^\mathrm{dis,min}_{i,a}$ and $P^\mathrm{dis,max}_{i,a}$ are drawn randomly also from $\left[0.1,0.6\right]$ and $\left[1.1,3.3\right]$ respectively. The charging and discharging efficiencies $\eta^{\mathrm{ch}}_{i,a}$ and $\eta^{\mathrm{dis}}_{i,a}$ are set to $0.87$ and $0.9$ respectively for EVs and to $0.91$ and $0.95$ respectively for batteries. Moreover, for batteries, $e_{i,a}^{\mathrm{SoC,ini}}$ and $e_{i,a}^{\mathrm{SoC,final}}$ are both set to $0.3e_{i,a}^{\mathrm{SoC,max}}$, whereas for EVs, which are required to be fully charged by $\tau^{\mathrm{end}}_{i,a \in \mathcal{A}_{i,4}}$, $e_{i,a}^{\mathrm{SoC,final}}$ is set equal to $e_{i,a}^{\mathrm{SoC,max}}$. The initial state of energy $e_{i,a}^{\mathrm{SoC,ini}}$ for an EV is set to $0.4e_{i,a}^{\mathrm{SoC,max}}$. Also, it is assumed that EVs are mostly required to be charged somewhere between $\tau^{\mathrm{start}}_{i,a \in \mathcal{A}_{i,4}}=7$pm and $\tau^{\mathrm{end}}_{i,a \in \mathcal{A}_{i,4}}=7$am. 

For Types 6 devices (air conditioners), the data for parameters $\psi_{i,a}$ and $\zeta_{i,a}$ is obtained by running the model initialization of the thermal model of a house in \cite{householdthermalmodel} for 10 distinct households with distinct geometries and thermal properties of the house materials. As for $P^\mathrm{th,min}_{i,a}$ and $P^\mathrm{th,max}_{i,a}$, their values are selected randomly from $\left[0.1,1\right]$ and $\left[2,5\right]$ respectively. The comfortable temperature range $\left[T_{i}^{\mathrm{in,min}},T_{i}^{\mathrm{in,max}}\right]$ is assumed to be $\left[\SI{18}{\degreeCelsius},\SI{25}{\degreeCelsius}\right]$ with the most comfortable temperature $T_{i}^{\mathrm{comf}}$ as $\SI{22.5}{\degreeCelsius}$. When it comes to $\tau_{i,a}$, the agents are divided into two groups. The first group consists of agents that prefer their air conditioners to be `on' from midday until late afternoon hours. The second group consists of agents preferring their air conditioners to be on from early evening hours untill midnight. Furthermore, the dissatisfaction parameter $\gamma^{\mathrm{th}}_{i,a}$ for Type 6 devices is selected randomly from $\left[0.001,0.15\right]$. Finally, the predicted input power $P_{i}^{\mathrm{PV},t}$ from PV panels is obtained from \cite{SHEMSwithEVandESS} and scaled by a factor drawn randomly from $\left[0.8,1.5\right]$.

In an effort to reflect realism (and break symmetry), it is assumed that only $40\%$ of the households have PV and battery storage systems and not more than $60\%$ of the households have EVs. It is also assumed that not more than $70\%$ of households have air conditioners. The coefficient of the quadratic cost component $c2^{t}$ is set to 0.007 from 8am to 2pm, 0.004 from 2pm to 7pm, 0.01 from 7pm to 12am, 0.003 from 12am to 5am, and 0.004 from 5am to 8am (as in \cite{FaithfulMDinDR}).

In all simulations, AMPL \cite{AMPL} is used as a frontend modeling language for the optimization problems along with Gurobi 6.0.5 \cite{gurobi} as a backend solver. Algorithm~\ref{algorithm} is coded in MATLAB and the interfacing between AMPL and MATLAB is made possible by AMPL's application programming interface. The simulations are all carried out on an Intel Core i7, 3.70GHz, 64-bit, 128GB RAM computing platform. Finally, the data set of 10 distinct households is replicated accordingly to generate the data sets for larger systems. 

\subsection{Centralized computation}\label{sec:centralized}
As a benchmark for comparison, the solution of the centralized problem \eqref{eq:mincost} along with its associated root-node gap are shown in Table~\ref{minimumcosts} for different system sizes. Table~\ref{minimumcosts} lists the total number of variables (Var), the number of binary variables (Bvar), the total number of constraints (Const), the solution to problem \eqref{eq:mincost} ($\mathcal{P}^{*}$) and its root-note gap (Gap (\%) ).
\begin{table}[!t]
%\normalsize
\centering
\caption{Solution of the centralized problem and its root-node gap.}
%\resizebox{\linewidth}{!}{%
\begin{tabular}{ r | c  c  c  c  c }
%\cline{7-8}
%\multicolumn{6}{c|}{} & \multicolumn{2}{c|}{Run-time (s)} \\\hline
\hline
$I$	&	Var	&	Bvar	&	Const	&	$\mathcal{P}^{*}$ (\$)	&	Gap (\%)	\\\hline
10	&	3734	&	1423	&	4205	&	11.25	&	39.13	\\\hline
20	&	7276	&	2774	&	8218	&	33.18	&	27.78	\\\hline
40	&	14360	&	5476	&	16244	&	104.44	&	19.28	\\\hline
80	&	28528	&	10880	&	32296	&	346.12	&	13.53	\\\hline
160	&	56864	&	21688	&	64400	&	1184.65	&	10.23	\\\hline
320	&	113536	&	43304	&	128608	&	4202.81	&	8.68	\\\hline
640	&	226880	&	86536	&	257024	&	15495.29	&	8.00	\\\hline
1280	&	453568	&	173000	&	513856	&	58857.48	&	7.70	\\\hline
2560	&	906944	&	345928	&	1027520	&	228738.14	&	7.72	\\\hline
\end{tabular}
\label{minimumcosts}
\end{table}

\begin{figure}[t]
\centering{
\psfrag{I}{\footnotesize $I$ \normalsize}
\psfrag{Computation time}{\footnotesize Computation time ($\SI{}{\second}$) \normalsize}
\psfrag{Centralized (MIQP)}{\footnotesize Centralized (MIQP) \normalsize}
\psfrag{Centralized (Relaxed)}{\footnotesize Centralized (Relaxed) \normalsize}
\psfrag{Distributed}{\footnotesize Distributed \normalsize}
\includegraphics[width=90mm] {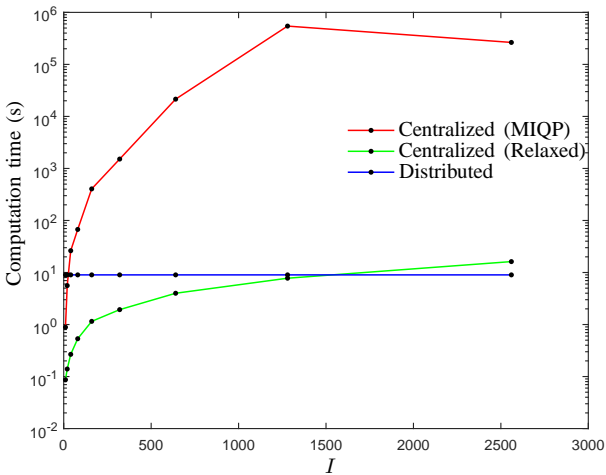}}
\caption{Scaling of the DR aggregation problem (semilogy plot).}
\label{computationtimes}
\end{figure}

As shown by Table~\ref{minimumcosts}, a peculiar attribute of problem \eqref{eq:mincost} is that it has a large root-node gap (loose relaxation). Problems that have a large root-node gap are, in practice, particularly hard to solve because they cast a heavier burden on the branch-and-cut algorithms, which manifests in longer computation times to reach optimality. The solver run-time of both the original nonconvex problem in \eqref{eq:mincost} and its convex relaxation are shown in Figure~\ref{computationtimes}. In fact, Gurobi's parameters are changed from their default values to ones that implement aggressive cuts generation (clique cuts, cover cuts and other Gurobi specific cuts) and aggressive presolve. Also, the primal simplex algorithm was chosen, instead of the default dual simplex algorithm, to solve the root-node and the node relaxations. This solver parameter tuning results in significant computation speedups for this specific problem, with up to $30\%$ faster computations in some instances. However, even with this solver parameter tuning, the solver run-time for the $I=1280$ and $I=2560$ test systems is more than 3 days, which is 3 times longer than the decision time horizon. Consequently, from a computational point of view, a centralized approach for solving the DR problem in this paper is not even suitable for day-ahead market clearing applications. 
%However, even with this solver parameter tuning, for the systems marked with $\dagger$ in Table~\ref{minimumcosts}, the solver failed to return the optimal solution within (more than) 168 hours at the default value of $10^{-4}$ of the MIP gap parameter. Therefore, for the systems marked with $\dagger$, the MIP gap parameter was reduced to $10^{-2}$ to obtain feasible near-optimal solutions within the computation times listed in Table~\ref{minimumcosts}.
On the other hand, the subproblems $\mathcal{D}_{i,\mu^{k},\nu^{k}} \left(\boldsymbol{\hat{\lambda}}_{k}\right)$, also being MIQPs, are easily handled by Gurobi in its default settings. 

\subsection{Distributed computation}\label{sec:distributed}
Algorithm~\ref{algorithm} is initialized with $\boldsymbol{\lambda}_{1}=\boldsymbol{0}$, $\kappa^{\mathrm{min}}=10^{-5}$, $\kappa^{1}=50$, $\alpha^{1}=8 \times 10^{-4}$, $\text{maxiterI}=30$, $\rho=0.3$, $\sigma=2$ and $\text{maxiterII}=30$. Consequently, the only parameter that requires tuning depending on system size is $\alpha^{\mathrm{min}}$. Obviously, this parameter has to be small enough so that the solution of the modified (double smoothed) dual function $\mathcal{D}_{\mu,\nu,\kappa} \left(\boldsymbol{\lambda}^{*}\right)$ is as close as possible to the original dual function $\mathcal{D} \left(\boldsymbol{\lambda}^{*}\right)$.

\begin{figure}[t]
\centering{
\psfrag{Iterations (k)}{\footnotesize Iterations (k) \normalsize}
\psfrag{Primal}{\footnotesize Primal \normalsize}
\psfrag{Dual}{\footnotesize Dual \normalsize}
\psfrag{Phase I}{\footnotesize \textcolor{red}{Phase I} \normalsize}
\psfrag{Phase II}{\footnotesize \textcolor{red}{Phase II} \normalsize}
\psfrag{Primal and Dual values}{\footnotesize $\mathcal{P}_{r}^{k}$ and $\mathcal{D}_{\mu^{k},\nu^{k},\kappa^{k}}(\hat{\boldsymbol{\lambda}}_{k})$ \normalsize}
%\psfrag{Primal and Dual values}{\footnotesize Primal and Dual objectives \normalsize}
\includegraphics[width=95mm] {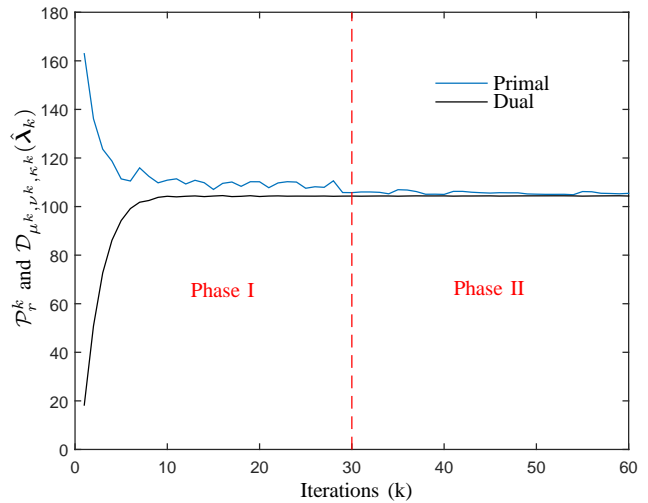}}
\caption{Evolution of the recovered primal and modified dual objectives for $I=40$.}
\label{40users0storage}
\end{figure}

\begin{figure}[t]
\centering{
\psfrag{Iterations (k)}{\footnotesize Iterations (k) \normalsize}
\psfrag{Primal}{\footnotesize Primal \normalsize}
\psfrag{Dual}{\footnotesize Dual \normalsize}
\psfrag{Phase I}{\footnotesize \textcolor{red}{Phase I} \normalsize}
\psfrag{Phase II}{\footnotesize \textcolor{red}{Phase II} \normalsize}
\psfrag{Primal and Dual values}{\footnotesize $\mathcal{P}_{r}^{k}$ and $\mathcal{D}_{\mu^{k},\nu^{k},\kappa^{k}}(\hat{\boldsymbol{\lambda}}_{k})$ \normalsize}
%\psfrag{Primal and Dual values}{\footnotesize Primal and Dual objectives \normalsize}
\includegraphics[width=95mm] {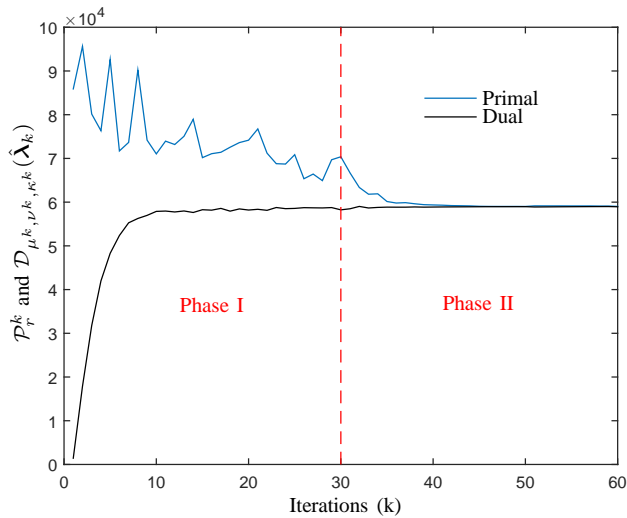}}
\caption{Evolution of the recovered primal and modified dual objectives for $I=1280$.}
\label{1280users0storage}
\end{figure}

The evolutions of the recovered primal iterates $\mathcal{P}_{r}^{k}$ and the double smoothed dual function $\mathcal{D}_{\mu^{k},\nu^{k},\kappa^{k}}(\hat{\boldsymbol{\lambda}}_{k})$ are displayed in Figure~\ref{40users0storage} and Figure~\ref{1280users0storage} for $I=40$ and $I=1280$ respectively. Figures~\ref{40users0storage} and~\ref{1280users0storage} show that $\mathcal{D}_{\mu^{k},\nu^{k},\kappa^{k}}(\hat{\boldsymbol{\lambda}}_{k})$ exhibits a fast and smooth convergence in less than $60$ iterations. In fact, $\mathcal{D}_{\mu^{k},\nu^{k},\kappa^{k}}\left(\hat{\boldsymbol{\lambda}}_{k}\right)-\mathcal{D}_{\mu^{k-1},\nu^{k-1},\kappa^{k-1}}\left(\hat{\boldsymbol{\lambda}}_{k-1}\right)$ decreases to less than $0.001$ in around $25$ iterations but the algorithm is kept running until $60$ iterations to allow $\left\|\nabla \mathcal{D}_{\mu^{k},\nu^{k},\kappa^{k}}\left(\hat{\boldsymbol{\lambda}}_{k}\right)\right\|$ to decrease enough to guarantee recovering a good primal solution. Furthermore, the convergence behavior of the dual iterates witnessed in Figures~\ref{40users0storage} and~\ref{1280users0storage} carries over to all the other problem instances. This feature is of paramount importance for the scalability of the algorithm. 

The algorithm has been found to suitably converge within $60$ iterations across a large number numerical tests on a vast array of test systems with different mixtures of devices and appliances. The algorithm can be terminated in less than $60$ iterations but this might come at the price of a lower quality solution; or it may require instance-specific parameter tuning to achieve the same quality solution in fewer iterations. Conversely, increasing the number of maximum iterations above $60$ can result in smaller optimality gaps but the marginal decrease in the recovered primal values is not high enough to warrant this increase. In summary, $60$ iterations has been found empirically to strike a good tradeoff between having a small number of iterations and recovering high quality feasible solutions. 

For all test systems, the optimality gap is measured as follows:
\begin{align*}
	\mathrm{Gap}=\frac{\left(\mathcal{P}_{r}^{S}-\mathcal{P}^{*}\right)}{\mathcal{P}^{*}} \times 100.
\end{align*}
The optimality gaps for the studied test systems are listed in Table~\ref{distributedresults}. Table~\ref{distributedresults} shows that the optimality gap does not exceed $0.48\%$ in all the considered test cases. In fact, further tuning parameters $\alpha^{1}$, $\alpha^{\mathrm{min}}$, $\rho$ and $\sigma$ can result in optimality gaps as low as $0.04\%$ but these results are not displayed here for the sake of keeping the algorithm as generic as possible. 
%The numbers that are marked with $\ddagger$ correspond to the test cases marked with $\dagger$ in Table~\ref{minimumcosts}. For these test systems, $\mathcal{P}^{*}$ is not the global optimum (see Section~\ref{sec:centralized}) and therefore the solution found by Algorithm~\ref{algorithm} might be better than $\mathcal{P}^{*}$ for the MIP gap parameter value of $10^{-2}$.

\begin{table}[!t]
%\normalsize
\centering
\caption{Algorithm parameter setting and optimality gap.}
%\resizebox{\linewidth}{!}{%
\begin{tabular}{ r | c  c  c  c  c }
\hline
$I$&$\mu^{\mathrm{min}}$&$S$&$\mathcal{P}_{r}^{S} (\$)$&$\mathcal{P}^{*} (\$)$& Opt. Gap (\%) \\\hline
10	&	$5 \times 10^{-6}$	&	54	&	11.26	&	11.25	&	0.11	\\\hline
20	&	$5 \times 10^{-6}$	&	53	&	33.25	&	33.18	&	0.22	\\\hline
40	&	$5 \times 10^{-6}$	&	41	&	104.87	&	104.44	&	0.41	\\\hline
80	&	$5 \times 10^{-6}$	&	43	&	347.73	&	346.12	&	0.46	\\\hline
160	&	$5 \times 10^{-6}$	&	60	&	1187.47	&	1184.65	&	0.24	\\\hline
320	&	$5 \times 10^{-6}$	&	60	&	4219.40	&	4202.81	&	0.39	\\\hline
640	&	$5 \times 10^{-6}$	&	56	&	15570.22	&	15495.29	&	0.48	\\\hline
1280	&	$5 \times 10^{-5}$	&	50	&	59013.32	&	58857.48	&	0.26	\\\hline
2560	&	$5 \times 10^{-5}$	&	59	&	229560.12	&	228738.14	&	0.36	\\\hline
\end{tabular}
\label{distributedresults}
\end{table}

\subsection{Discussion}\label{sec:Discussion}
\begin{figure}[t]
\centering{
\psfrag{Iterations (k)}{\footnotesize Iterations (k) \normalsize}
\psfrag{Primal}{\footnotesize Primal \normalsize}
\psfrag{Dual}{\footnotesize Dual \normalsize}
\psfrag{Primal and Dual values}{\footnotesize $\mathcal{P}_{r}^{k}$ and $\mathcal{D}_{\mu^{k},\nu^{k},\kappa^{k}}(\hat{\boldsymbol{\lambda}}_{k})$ \normalsize}
%\psfrag{Primal and Dual values}{\footnotesize Primal and Dual objectives \normalsize}
\includegraphics[width=95mm] {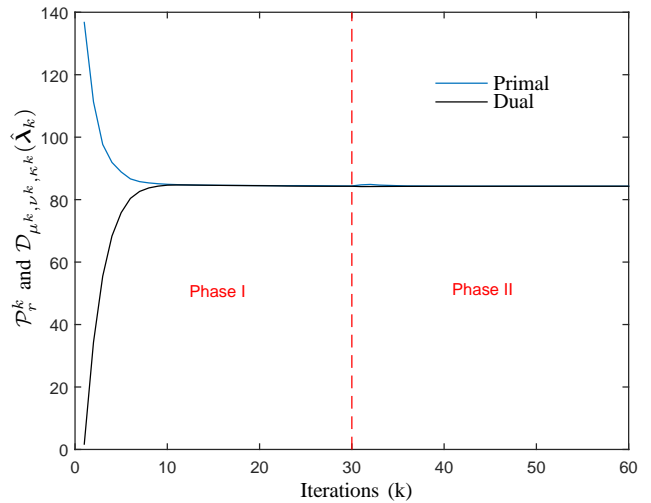}}
\caption{Evolution of the recovered primal and modified dual objectives of the relaxed (convex) version of the test system with $I=40$.}
\label{40relaxed}
\end{figure}

A common trait for all the test systems is the oscillation of the recovered primal iterates. These oscillations stem from a combination of two reasons. The first reason is that because the households have mixed-integer variables, their feasible scheduling sets are disjoint. Therefore, a change in the price signal $\boldsymbol{\hat{\lambda}}_{k}$ can result in $x^{t}_{i,\mu^{k},\nu^{k},\boldsymbol{\hat{\lambda}}_{k}}$ changing in discrete steps. The second reason is that the test systems are a replication of the data set of 10 distinct households. Therefore all the similar households will exhibit the same best response to the price signal $\boldsymbol{\hat{\lambda}}_{k}$ which entails that the effect of the first reason will be magnified on the collective level. By this reasoning, this oscillatory behavior of the recovered primal iterates should not exist when all the variables are continuous and the problem is convex. Indeed, Figure~\ref{40relaxed} shows that the recovered primal iterates of a relaxed version of the test system with $I=40$ do not exhibit this oscillatory behavior. It is also evident from Figure~\ref{40relaxed} that an optimal solution can be found within $60$ iterations for a relaxed version of the adopted DR problem. In fact, a close inspection of Figure~\ref{40relaxed} and Figure~\ref{40users0storage} shows that the modified dual function exhibits a very similar convergence behavior in both the original nonconvex problem and its convex relaxation.

\begin{figure}[t]
\centering{
\psfrag{Iterations (k)}{\footnotesize Iterations (k) \normalsize}
\psfrag{Primal}{\footnotesize Primal \normalsize}
\psfrag{Dual}{\footnotesize Dual \normalsize}
\psfrag{Primal and Dual values}{\footnotesize $\mathcal{P}_{r}^{k}$ and $\mathcal{D}(\boldsymbol{\lambda}_{k})$ \normalsize}
\includegraphics[width=95mm] {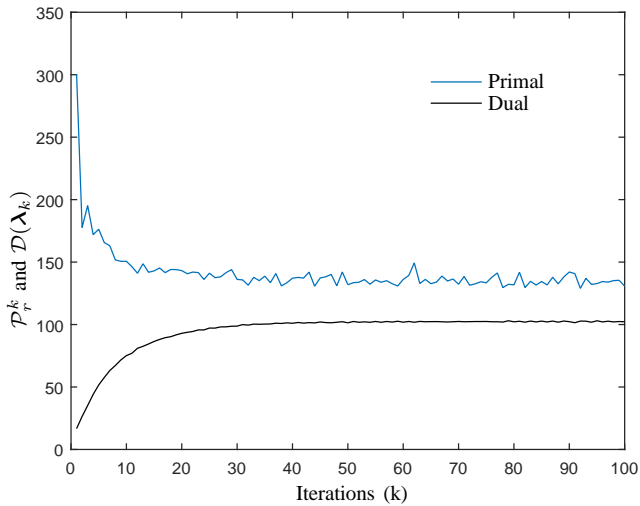}}
\caption{Evolution of the recovered primal and original dual objectives of the test system with $I=40$ using a conventional gradient method with a step size of $0.0005$.}
\label{40LR}
\end{figure}

Moreover, to see the superiority of the fast gradient algorithm (applied to the double smoothed dual function) over a conventional gradient method (applied to the original nonsmooth dual function), Figure~\ref{40LR} shows the evolution of the recovered primal and original dual objectives of the test system with $I=40$ using a conventional gradient method with a step size of $0.0005$. In this case, the dual function is nonsmooth and exhibits an oscillatory behavior\footnote{Refer to Figure~\ref{dualplot} for a geometric interpretation of the oscillations.} just like the recovered primal. This has an adverse effect on the stopping criteria of the algorithm. Additionally, it is clear from Figure~\ref{40LR} that the duality gap (and also the optimality gap) does not decrease below $15\%$. The same observation applies to all the considered test systems.

Finally, the computational effort of the proposed fast gradient method is distributed among the agents. The MIQPs solved by agents $i \neq 0$ take less than $0.15$ seconds to solve in the worst case and the aggregator subproblem and primal recovery problems\footnote{The aggregator subproblem and the primal recovery problem are both convex QPs.} require less than $0.01$ seconds each to solve in the worst case. Therefore, as shown in Figure~\ref{computationtimes}, the parallel solve time of the algorithm is at most $0.15 \times 60=9$ seconds (neglecting communication overhead), which can be ideal for an on-line version of this problem. In a practical implementation, the convergence time of the  algorithm is expected to align with the local energy market's operation. For example, in the Australian National Electricity Market, supply procurement auctions are run for each $30$ minute period. In this case, $30$ minutes is ample time compared to the $9$ seconds required for the proposed algorithm to complete $60$ iterations and find a near-optimal solution.

\section{Conclusion}\label{sec:conclusion}
The aim of this work is to demonstrate the scalability of a fast gradient algorithm applied to the double smoothed dual function of a large-scale nonconvex DR problem comprising expressive household models and mixed-integer variables. This work demonstrates how to recover a near-optimal solution within a preset small number of iterations and minimal parameter tuning. More specifically, the solutions recovered from the algorithm are on average within $0.32\%$ of the optimum. Additionally, the results show that the convergence of the algorithm exhibits a similar behavior across the studied test systems, which corroborates the method's scalability. The paper also provides a geometrical insight into the dual problem of the adopted nonconvex DR model and highlights the inefficacy of the conventional gradient method in solving this specific problem.

The work in this paper can be extended in several directions. The formulations in this paper can be easily extended to account for reverse power flow constraints, aggregator level renewable energy resources and aggregator controlled storage. Additionally, the DR problem in this paper can be extended to incorporate the nonlinear characteristics of hot-water systems, fuel cells, micro-CHP and a second order thermal model of a household. In this case, the resulting DR problem is a MINLP that has a nonconvex relaxation. Finally, future work will involve extending the DR problem in this paper to account for the underlying power distribution network through incorporating AC power flow and system operational constraints.

\section{Acknowledgment}\label{sec:acknowledgment}
This research was partly supported by Ausgrid and the Australian Research Council under Australian Research Council's Linkage Projects funding scheme (project number LP110200784).

\bibliographystyle{IEEEtran}
{\footnotesize
\bibliography{ODMcitations}}

\begin{IEEEbiography}[{\includegraphics[width=1in,height=1.25in,clip,keepaspectratio]{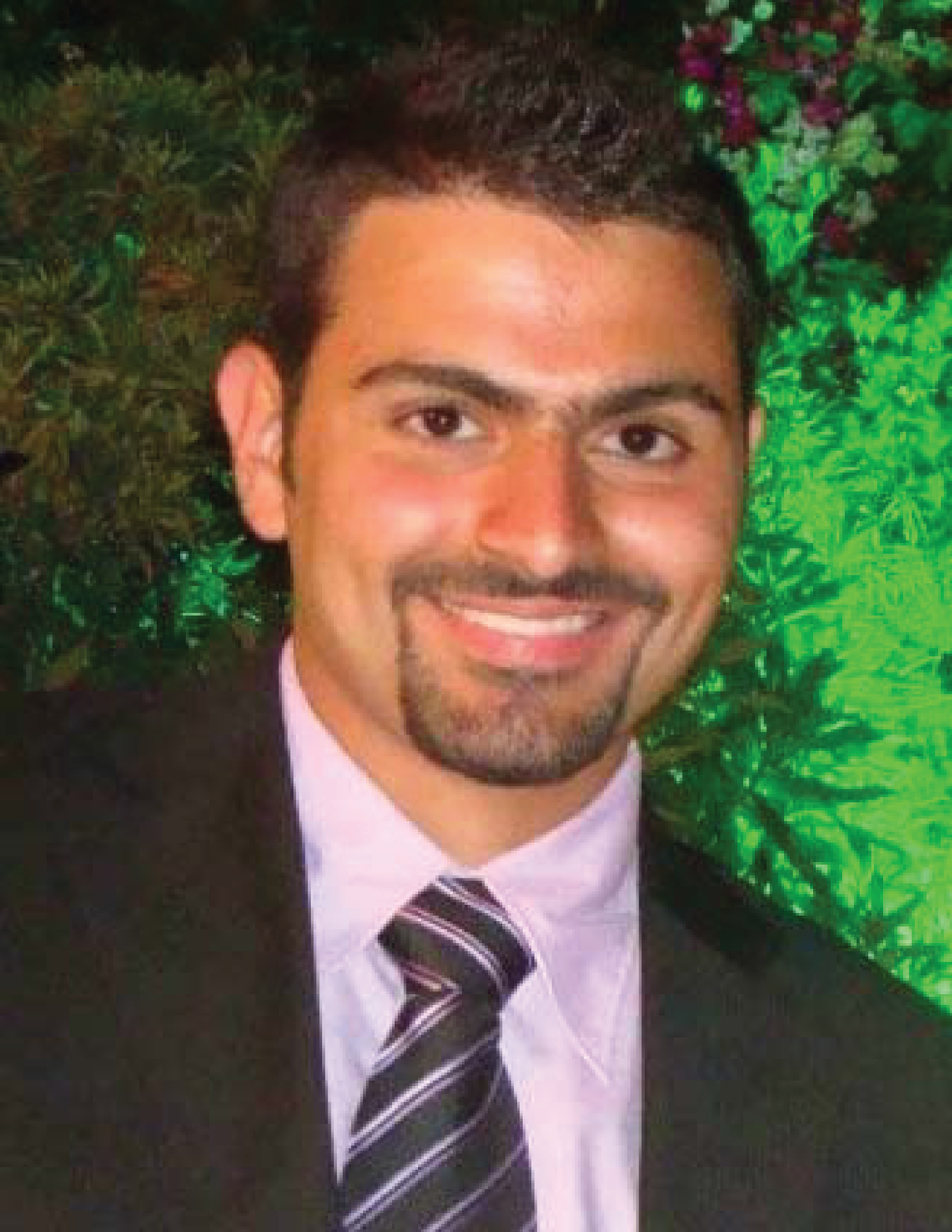}}]{Sleiman Mhanna}
(S'13) received the B.Eng. degree (with high distinction) from the Notre Dame University, Lebanon, and the M.Eng. degree from the American University of Beirut, Beirut, Lebanon, in 2010 and 2012, respectively, both in electrical engineering. He is currently pursing the Ph.D. degree with the School of Electrical and Information Engineering, Centre for Future Energy Networks, University of Sydney, Sydney, NSW, Australia.

His research interests include distributed methods and game theoretic analysis in power systems, demand response pricing mechanisms, and grid integration of distributed energy resources.
\end{IEEEbiography}

\begin{IEEEbiography}[{\includegraphics[width=1in,height=1.25in,clip,keepaspectratio]{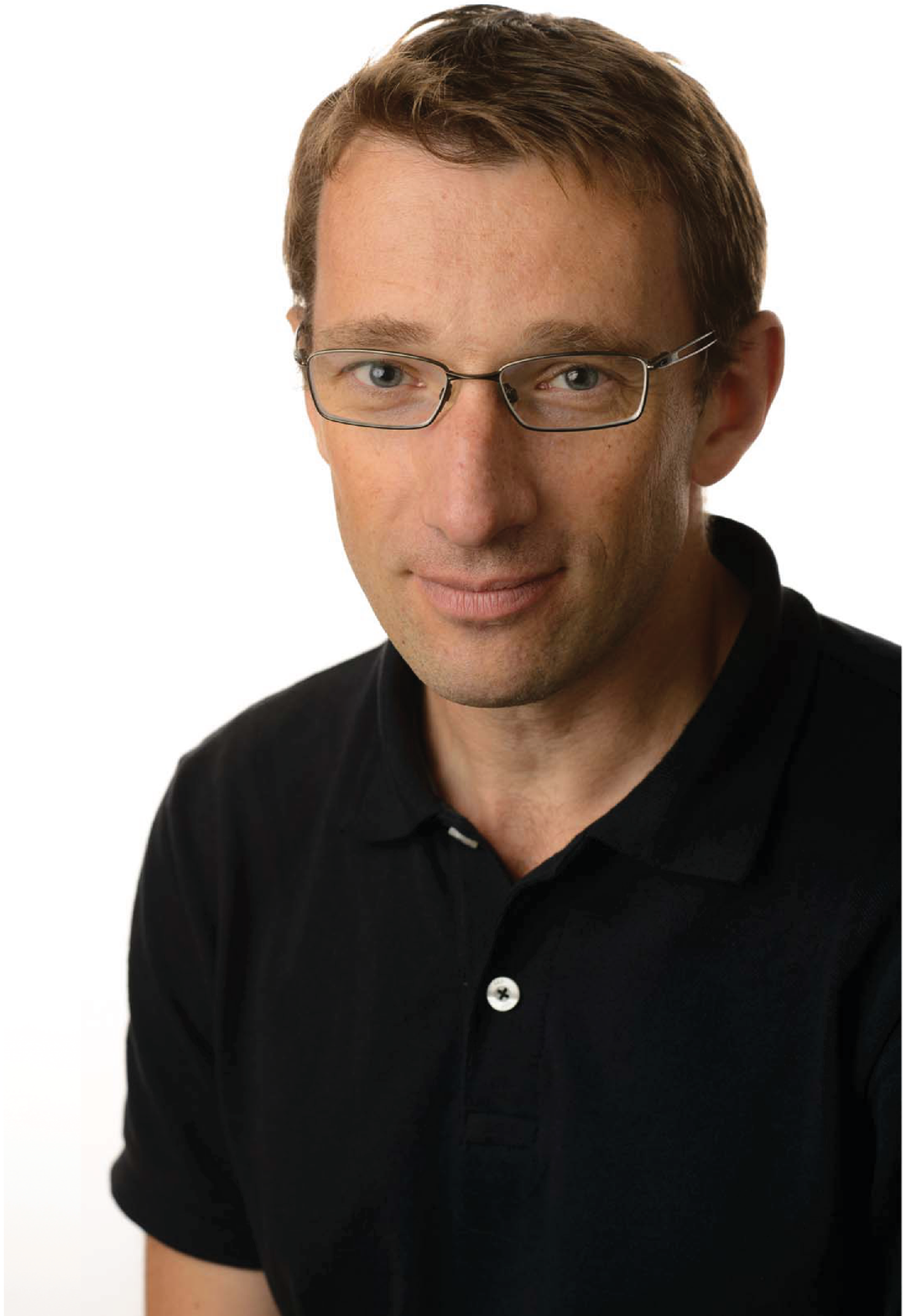}}]{Gregor Verbi\v{c}}
(S'98--M'03--SM'10) received the B.Sc., M.Sc., and Ph.D. degrees in electrical engineering from the University of Ljubljana, Ljubljana, Slovenia, in 1995, 2000, and 2003, respectively. In 2005, he was a North Atlantic Treaty Organization-Natural Sciences and Engineering Research Council of Canada Post-Doctoral Fellow with the University of Waterloo, Waterloo, ON, Canada. Since 2010, he has been with the School of Electrical and Information Engineering, University of Sydney, Sydney, NSW, Australia. His expertise is in power system operation, stability and control, and electricity markets.

His current research interests include integration of renewable energies into power systems and markets, optimization and control of distributed energy resources, demand response, and energy management in residential buildings. Dr. Verbi\v{c} was a recipient of the IEEE Power and Energy Society Prize Paper Award in 2006. He is an Associate Editor of the IEEE TRANSACTIONS ON SMART GRID.
\end{IEEEbiography}

\begin{IEEEbiography}[{\includegraphics[width=1in,height=1.25in,clip,keepaspectratio]{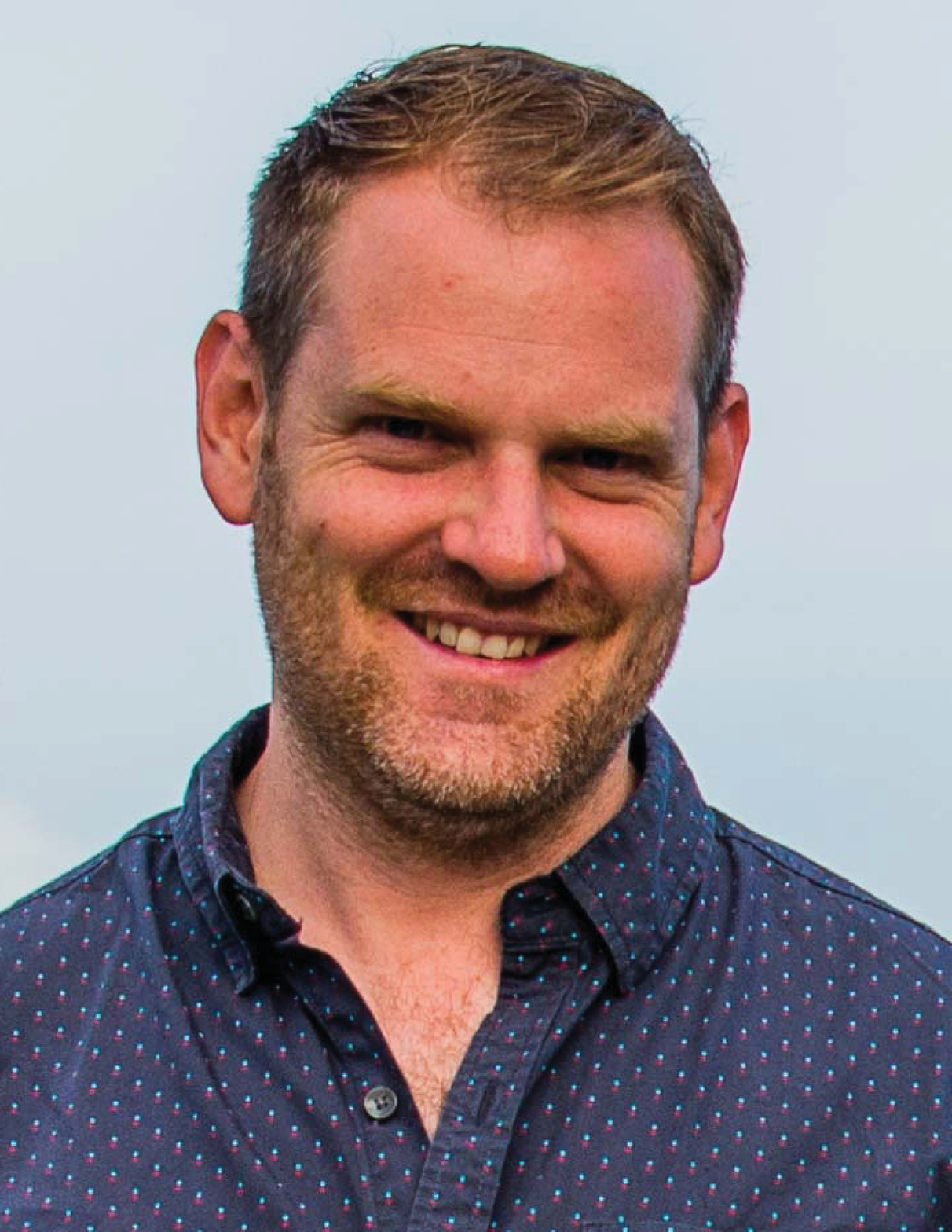}}]{Archie C. Chapman}
(M'14) received the B.A. degree in math and political science, and the B.Econ. (Hons.) degree from the University of Queensland, Brisbane, QLD, Australia, in 2003 and 2004, respectively, and the Ph.D. degree in computer science from the University of Southampton, Southampton, U.K., in 2009.

He is currently a Research Fellow in Smart Grids with the School of Electrical and Information Engineering, Centre for Future Energy Networks, University of Sydney, Sydney, NSW, Australia. He has experience in game-theoretic and reinforcement learning techniques for optimization and control in large distributed systems.
\end{IEEEbiography}

\end{document}